\documentclass[11pt]{article}
\usepackage{epsfig}
\usepackage{graphicx}
\usepackage{citesort}
\usepackage{amsmath,amssymb,amsfonts}
\usepackage{amsmath}
\usepackage{amsfonts}
\usepackage{amssymb}
\usepackage{dsfont,amsthm}
\usepackage[dvips]{feynmp}
\usepackage{slashed}
\usepackage{graphicx,ifpdf}
\usepackage{upgreek}
\usepackage{floatflt}
\usepackage{amsxtra,mathrsfs,latexsym}\usepackage{dsfont}
\newcommand{\diff}{\text{\rm d}}

\newcommand{\euler}{\text{\rm e}}\newcommand{\imu}{\text{\rm i}}

\newcommand{\field}[1]{\mathds{#1}}

\newcommand{\Z}{\field {Z}}

\ifpdf
\fi
\usepackage[dvips,textwidth=16.0cm,textheight=24.7cm]{geometry}
\numberwithin{equation}{section}

\newcommand{\beq}{\begin{equation}}
\newcommand{\eeq}{\end{equation}}
\newcommand{\bea}{\begin{eqnarray}}
\newcommand{\eea}{\end{eqnarray}}

\begin{document}
\title{\bf Thermodynamics of a three-flavor\\nonlocal  Polyakov--Nambu--Jona-Lasinio model \footnote{Work supported in part by BMBF, GSI, the DFG Excellence Cluster ``Origin and Structure of the Universe'' and by the Elitenetzwerk Bayern.}}
\author{
T.~Hell, S.~R\"o{\ss}ner, M.~Cristoforetti, and W.~Weise\\
\\{\small Physik-Department, Technische Universit\"at M\"unchen, D-85747 
Garching, Germany}
}
\date {\today}
\maketitle

\begin{abstract}

The present work generalizes a nonlocal version of the Polyakov-loop-extended Nambu and Jona-Lasinio (PNJL) model to the case of three active quark flavors, with inclusion of the axial $\text{U}(1)$ anomaly. 
Gluon dynamics is incorporated through a gluonic background field, expressed in terms of the Polyakov loop. The thermodynamics of the nonlocal PNJL model accounts for both chiral and deconfinement transitions. Our results obtained in mean-field approximation are compared to lattice QCD results for $N_\text{f}=2+1$ quark flavors. Additional pionic and kaonic contributions to the pressure are calculated in random phase approximation. Finally, this nonlocal 3-flavor PNJL model is applied to the finite density region of the QCD phase diagram. It is confirmed that the existence and location of a critical point in this phase diagram depend sensitively on the strength of the axial $\text{U}(1)$ breaking interaction.

\end{abstract}

\begin{section}{Introduction}

Investigating the phase diagram of strongly interacting matter is a persistently challenging theme of nuclear and high-energy physics. Ultrarelativistic heavy ion collisions at the Relativistic Heavy Ion Collider give evidence that a strongly correlated quark-gluon phase is formed at temperatures above $200\,\text{MeV}$, in accordance with lattice QCD calculations (at zero baryon chemical potential) which suggest a transition temperature in the same range for chiral symmetry restoration and confinement-deconfinement transitions. The ALICE experiment at the Large Hadron Collider  will shed further light on the region of higher temperatures and low quark chemical potentials.

Lattice QCD computations can so far not be extended systematically into the region of larger quark chemical potentials $\mu$, but expansions around $\mu=0$ do suggest the existence of a critical point at which the chiral crossover transition at small $\mu$ turns into a first-order phase transition. The precise location of this critical point and possibly even its mere existence are still controversial. Given that, at the present exploratory stage, neither lattice QCD nor experiment can yet map out the QCD phase diagram over large areas in the $(T,\mu)$ plane, it is useful to work with models in order to identify the dynamically relevant degrees of freedom.

Models of the Nambu and Jona-Lasinio (NJL) type \cite{Nambu} have been quite useful for orientation in this context \cite{Hell,Scarpettini,Vogl,Pisarski,Meisinger,Fukushima1,Fukushima2,Ratti1,Simon1,Ratti2,Simon2,Blaschke,Sasaki,Abuki,Fukushima3,Buballa}  as they properly incorporate the chiral symmetry breaking scenario of low-energy QCD. A basic element of such models is the gap equation connecting the chiral condensate and the dynamical quark mass, providing a mechanism for spontaneous chiral symmetry breaking and the generation of quark quasiparticle masses. Thermodynamic aspects of confinement, while absent in the original NJL model, can be implemented by a synthesis with Polyakov-loop dynamics. The resulting Polyakov--Nambu--Jona-Lasinio PNJL model \cite{Fukushima1,Fukushima2,Ratti1,Simon1,Ratti2,Simon2,Blaschke,Sasaki,Abuki,Buballa} has been remarkably successful in describing the two-flavor thermodynamics of QCD. However, this earlier version of the PNJL approach still worked with an artificial momentum space cutoff, $\varLambda_\text{NJL}\approx (0.6\text{--}0.7)\,\text{GeV}$, which prohibits establishing connections with well-known properties of QCD at higher momentum scales such as the running coupling and momentum-dependent quark mass function. Furthermore, thermodynamically consistent results at high temperatures and densities cannot be achieved using the original (local) PNJL model. In particular, a meaningful extrapolation to the high-density region with its variety of color-superconducting phases cannot be performed once the quark Fermi momentum becomes comparable to the NJL cutoff. The nonlocal PNJL model does not have such \emph{a priori} limitations.

In fact the nonlocal two-flavor PNJL model \cite{Hell} solves this problem by introducing momentum-dependent quark interactions that permit realizing the high-momentum interface with QCD and Dyson-Schwinger calculations at the level of the quark quasiparticle propagators. The present work takes a next major step by extending this nonlocal PNJL model to $N_\text{f}=3$ flavors, now incorporating the strange quark. This step involves a detailed study of the axial $\text{U}(1)$ anomaly, its role in separating the flavor singlet component of the pseudoscalar meson nonet from the Nambu-Goldstone boson sector, and its thermodynamical implications.  It will turn out, that the nonlocal PNJL model does not suffer from the pathologies mentioned above, and hence it does not have any \emph{a priori} limitations. Therefore, it is well-suited to investigate the high-density and high-temperature region of strongly interacting matter.

From hadron spectroscopy it is well-known that only eight of the nine lightest pseudoscalar mesons  (the pions, the kaons, and the eta meson) have pseudo-Goldstone boson character. The eta-prime meson, on the other hand, has a mass of $m_{\eta'}\simeq958\,\text{MeV}$, about $400\,\text{MeV}$ higher than the eta mass: the $\eta'$ meson is not part of the Nambu-Goldstone multiplet. The reason is the axial $\text{U}(1)_\text{A}$ anomaly \cite{Adler} and its realization in QCD. The flavor singlet axial current, $j_5^\mu(x)$, is not conserved. In fact one has $\partial_\mu j_5^\mu\propto \vec{E}_a\cdot\vec{B}^a$; i.\,e. the divergence of the singlet axial current is proportional to the product of color electric and magnetic gluon fields. In the interpretation given by 't~Hooft \cite{Hooft}, the axial anomaly is induced by instanton effects and translates into an axial $\text{U}(1)_\text{A}$ breaking effective interaction between quarks that has the form of an $N_\text{f}\times N_\text{f}$ determinant \cite{KM} of right- and left-handed quark bilinears, $\psi_i(1\pm\gamma_5)\psi_j$. For $N_\text{f}=3$ this is a genuine six-point vertex involving all three u, d and s quarks simultaneously. In its local form, this 't~Hooft determinant interaction has been widely used in previous three-flavor NJL model calculations \cite{Klimt1,Klimt2,Rehberg,Fukushima3,Buballa}. Our present work generalizes this $\text{U}(1)_\text{A}$ breaking interaction for its use as part of the nonlocal $N_\text{f}=3$ PNJL model. 

This paper is organized as follows. Sections \ref{model} and \ref{thermodynamics} introduce and develop the nonlocal PNJL model. Its connection with the instanton model is outlined. A comparison with results from Dyson-Schwinger calculations of the Landau gauge QCD is performed. Properties of the pseudoscalar mesons, such as masses and decay constants, are derived within a systematic expansion of the action around the mean-field limit. Fundamental low-energy theorems such as the Gell-Mann--Oakes--Renner relation are shown to hold within the nonlocal framework. Section~\ref{thermodynamics} proceeds with the thermodynamics of the nonlocal PNJL model using the Matsubara formalism. The Polyakov loop $\Phi$ is introduced as an (approximate) order parameter for the confinement-deconfinement transition. The quarks are coupled to $\Phi$ in the usual minimally gauge invariant way. The gap equations following from the PNJL effective action determine the temperature dependence of the chiral up-, down-, and strange-quark condensates and of the Polyakov loop. We also take into account contributions to the pressure beyond mean-field approximation, calculating pionic and kaonic quark-antiquark correlations in random phase approximation. The nonlocal PNJL model is then applied to the finite density case, introducing nonvanishing (quark) chemical potentials. We sketch a simplified version of the QCD phase diagram (without inclusion of diquark condensates). The variation of the ``critical point" with changing strength of the 't~Hooft interaction is investigated.  Section~\ref{summary} presents conclusions and an outlook.

\end{section}

\begin{section}{Nonlocal three-flavor Nambu--Jona-Lasinio model}\label{model}

This section extends the previously developed nonlocal NJL model \cite{Hell} to $N_\text{f}=3$ flavors. It is designed so as to correctly implement the spontaneously broken chiral $\text{SU}(3)_\text{R}\times\text{SU}(3)_\text{L}$ symmetry of QCD at zero temperature together with the anomalously broken axial $\text{U}(1)_\text{A}$ symmetry.

\begin{subsection}{Chirally invariant nonlocal action}

As in our previous work \cite{Hell}, the construction of the interaction part of the effective quark action in Euclidean space\footnote{If not stated otherwise, all quantities are expressed in Euclidean space-time, i.\,e. $x=(x_4,\vec{x}\,)=(\imu x_0,\vec{x}\,)$, $\gamma_4:=\imu\gamma_0$, etc.} is guided by the nonlocal coupling of color currents. A Fierz transformation leads to the following generic form of the nonlocal 4-point interaction,
\begin{equation}\label{sint4}
	\tilde{\mathcal{S}}_\text{int}^{(4)}=\sum_\alpha c_\alpha\int\diff^4 x\int\diff^4 z\,\bar\psi\left(x+\frac{z}{2}\right)\varGamma_\alpha\,\psi\left(x-\frac{z}{2}\right)\,\mathcal{G}(z)\,\bar\psi\left(x-\frac{z}{2}\right)\varGamma^\alpha\,\psi\left(x+\frac{z}{2}\right).
\end{equation}
The nonlocality distribution, $\mathcal{G}(z)$, is proportional to the gluonic field correlator. As in Ref.~\cite{Hell} we restrict ourselves to the diagonal part of the Lorentz tensor representing this correlator. The coefficients $c_\alpha$ result from a Fierz transform of the color current-current interaction on which the action \eqref{sint4} is based. The $\varGamma_\alpha$ are a set of Dirac, flavor and color matrices. Their combination is determined by the Fierz transform just mentioned. The quark field is given as  $\psi(x)=\left( u(x),d(x),s(x)\right)^\top$ in terms of the u-, d-, and s-quark fields. Their current quark masses are collected in the mass matrix $\hat m_q=\text{diag}(m_u,m_d,m_s)$.

The coupling term \eqref{sint4} of the action shares with full QCD a global chiral $\text{U}(3)_\text{R}\times\text{U}(3)_\text{L}$ symmetry. The axial $\text{U}(1)_\text{A}$ subgroup of $\text{U}(3)\times\text{U}(3)$ is broken by the anomaly which will be treated by an additional  term of the action, to be described in subsection~\ref{anomaly}. Apart from a strictly conserved $\text{U}(1)$ symmetry associated with baryon number, the remaining chiral $\text{SU}(3)_\text{R}\times\text{SU}(3)_\text{L}$ symmetry undergoes dynamical (spontaneous) breaking through $\tilde{\mathcal{S}}^{(4)}_\text{int}$, down to flavor $\text{SU}(3)$ (the well-known ``eightfold way''). Explicit chiral and flavor symmetry breaking corrections are then introduced by the quark mass term of the Lagrangian, $\bar\psi\,\hat m_q \psi$.

In Eq.~\eqref{sint4} we restrict ourselves to the color-singlet pair of scalar and pseudoscalar operators as a minimal $\text{U}(3)\times\text{U}(3)$ symmetric combination:
\begin{equation*}
 	\varGamma_{\alpha\in\{0,1,\dots,8\}}=(1,\imu\gamma_5)\times\lambda_\alpha,
\end{equation*}
with the Gell-Mann matrices $\{\lambda_0,\lambda_1,\dots,\lambda_8\}$ specifying a basis in flavor space, where we have introduced the singlet matrix $\lambda_0:=\sqrt{\frac{2}{3}}\,\text{diag}(1,1,1)$. Other less relevant operators (such as Lorentz vectors and axial vectors) will be ignored in the present work. (See however the investigation of the role of vector current interactions in Ref.~\cite{Maris}.)

Furthermore, we replace the dimensionful nonlocality distribution function $\mathcal{G}$ in Eq.\,\eqref{sint4} by a coupling strength $G/2$ (with mass dimension $-2$) and a normalized distribution $\mathcal{C}$, i.\,e.
\begin{equation}\label{nambufunction}
 	\mathcal{G}(z)=\dfrac{G}{2}\mathcal{C}(z),
\end{equation}
with $\int\diff^4 z\,\mathcal{C}(z)=1$. Note that the standard (local) NJL model follows for the limiting case $\mathcal{C}(z)=\delta^{(4)}(z)$, with the understanding that a momentum space cutoff is introduced in this local limit in order to regularize loop integrals.

\begin{subsubsection}{Schwinger-Dyson equation}

Following Ref.~\cite{Hell}, the next obvious step is a bosonization of the action $\tilde{\mathcal{S}}_\text{int}^{(4)}$. Such a procedure was performed, e.\,g., in Refs.~\cite{Praschifka,Cahill,Ball} leading to the following Schwinger-Dyson (SD) equation (gap equation) in momentum space for the dynamical quark masses $M_u(p)=M_d(p)=M_s(p)\equiv M(p)$ in the chiral limit (i.\,e., $m_u=m_d=m_s=0$):
\begin{equation}\label{sde}
	M(q)=8 N_\text{c}G\int\dfrac{\diff^4 p}{(2\pi)^4} \tilde{\mathcal{C}}(q-p)\dfrac{M(p)}{p^2+M^2(p)}\,.
\end{equation}
Here the Euclidean Fourier transform $\tilde{\mathcal{C}}(p)=\int\diff^4 z\,\euler^{-\imu p z}\mathcal{C}(z)$ of the distribution $\mathcal{C}(z)$ has been introduced, with $\tilde{\mathcal{C}}(p=0)=1$.
The mass function $M(p)$ is interpreted as the momentum-dependent dynamical (constituent) quark mass which, in the case of $M(p)\not\equiv0$, expresses the spontaneous chiral symmetry breaking. The gap equation~\eqref{sde} can be solved iteratively relying on Banach's fix-point theorem. It turns out indeed that such a procedure leads to a nontrivial solution with nonvanishing $M$ (see Fig.~\ref{DSEplot}).

\end{subsubsection}

\begin{subsubsection}{Four-fermion separable interaction}

The full solution of the SD equation \eqref{sde} is, however, not practical for our present purposes. If one were to solve such an integral equation for finite temperature one would be confronted with the full complexity of finite temperature SD calculations (cf.~Refs.~\cite{Roberts,Alkofer,Fischer}). We approximate the distribution $\mathcal{C}$ by a separable form, replacing
\begin{equation*}
 	\mathcal{G}(z)=\dfrac{G}{2}\,\mathcal{C}(z)\to\dfrac{G}{2}\int\diff^4 z'\,\mathcal{C}\left(\!z+\dfrac{z'}{2}\right)\mathcal{C}\left(\!z-\dfrac{z'}{2}\right)
\end{equation*}
(compare also Ref.~\cite{Burden}). The model using this separable ansatz will be worked out starting from the next subsection. A comparison between the full SD calculation and the separable approach  in Sec.~\ref{sdsepcomparison} demonstrates that the separable action is a very good approximation to the full SD-formalism for all practical purposes.
We can now write the chirally invariant four-fermion interaction in a more tractable form: 
\begin{equation}\label{sint4sep}
	\tilde{\mathcal{S}}_\text{int}^{(4)}\to \mathcal{S}_\text{int}^{(4)}=-\dfrac{G}{2}\int\diff^4 x\left[ j_\alpha^S(x)j_\alpha^S(x)+j_\alpha^P(x) j_\alpha^P(x)\right],
	\end{equation}
where $j_\alpha^S$ and $j_\alpha^P$ are scalar and pseudoscalar densities, given by
\begin{equation}\label{j}
	\begin{aligned}
		j_\alpha^S(x)&=\int\diff^4z\,\bar\psi\!\left(x+\frac{z}{2}\right)\,\mathcal{C}(z)\,\lambda_\alpha\psi\!\left(x-\frac{z}{2}\right),\\
		j_\alpha^P(x)&=\int\diff^4z\,\bar\psi\!\left(x+\frac{z}{2}\right)\,\mathcal{C}(z)\,\imu\gamma_5\lambda_\alpha\psi\!\left(x-\frac{z}{2}\right).
	\end{aligned}
\end{equation}
The particular functional form of $\mathcal{C}(z)$ used in this work will be given in Sec.~\ref{parameters} (see also Fig.~\ref{CvsCI}).

\end{subsubsection}

\end{subsection}

\begin{subsection}{Axial {\boldmath{$\text{U}(1)$}} anomaly and instantons}\label{anomaly}

So far, the generalization from the two-flavor to the three-flavor case has been straightforward. The more difficult task concerning the extension of the nonlocal NJL approach to three flavors is the implementation of the mechanism that breaks the $\text{U}(1)_\text{A}$ symmetry, leading to the relatively large mass of the $\eta'$ meson. 

At the level of an effective interaction between quarks, this can be accomplished by introducing an interaction term proportional to the Kobayashi-Maskawa-'t~Hooft determinant \cite{Hooft,KM},
\begin{equation*}
\int\diff^4 x\,\left[\det\mathcal{J}_+(x)+\det\mathcal{J}_-(x)\right]
\end{equation*}
with
\begin{equation}\label{thooft}
\left(\mathcal{J}_\pm(x)\right)_{ij}=\int\diff^4 z\,\bar\psi_j\!\left(x+\frac{z}{2}\right)\dfrac{1}{2}(1\mp\gamma_5)\,\mathcal{K}(z)\,\psi_i\!\left(x-\frac{z}{2}\right),
\end{equation}
where $\mathcal{K}(z)$ represents the distribution of the $\text{U}(1)_\text{A}$ breaking interaction strength. For $N_\text{f}=3$, this flavor determinant generates a genuine 3-body interaction (or 6-quark vertex) in which the u, d, and s quarks participate simultaneously.

In his original paper \cite{Hooft}, 't~Hooft derived the $\text{U}(1)_\text{A}$ breaking interaction starting from instantons. The instanton liquid model gives a simple expression for $\mathcal{K}$ as the density of zero modes. Its Fourier transform $\tilde{\mathcal{K}}(p)=\int\diff^4z\,\euler^{-\imu p z}\mathcal{K}(z)$ is written \cite{Schaefer,Cristoforetti} in terms of Bessel functions and a characteristic instanton size, $d\simeq0.35\,\text{fm}$, as follows:
\begin{equation}\label{instantonc}
	\tilde{\mathcal{K}}(p)=\pi p^2 d^2 \dfrac{\diff}{\diff \xi}\big[I_0(\xi)K_0(\xi)-I_1(\xi)K_1(\xi)\big]\qquad\text{with } \xi=\frac{|p| d}{2}\ .
\end{equation}

At this point we can anticipate a result of our present studies (see Fig.~\ref{CvsCI}), namely that the distribution $\mathcal{C}$ in the nonlocal chiral four-fermion interaction \eqref{sint4sep} turns out to be very close to the distribution $\mathcal{K}$ that characterizes the instanton induced interaction \eqref{thooft}. We can therefore set $\mathcal{C}=\mathcal{K}$ and work with a universal distribution involving a single scale, e.\,g., the instanton size $d$. It is then not surprising that the local version of the (classic) NJL model operates with a momentum cutoff scale that reflects the inverse instanton size, $\varLambda_\text{NJL}\sim1/d$. 

With these preparations and following the steps outlined in Appendix~\ref{hooftappendix}, the 't~Hooft determinant can be written in terms of the nonlocal scalar and pseudoscalar densities \eqref{j}, leading to the following six-fermion part of the action:
 \begin{equation}\label{sint6sep}
 \mathcal{S}_\text{int}^{(6)}=-\dfrac{H}{4}\int\diff^4 x\,\mathcal{A}_{\alpha\beta\gamma}\left[j_\alpha^S(x)j_\beta^S(x)j_\gamma^S(x)-3j_\alpha^S(x)j_\beta^P(x)j_\gamma^P(x)\right],
 \end{equation}
where the constants $\mathcal{A}_{\alpha\beta\gamma}$ are expressed in terms of the Gell-Mann matrices according to
\begin{equation*}	\mathcal{A}_{\alpha\beta\gamma}:=\dfrac{1}{3!}\varepsilon_{ijk}\varepsilon_{mn\ell}\left(\lambda_\alpha\right)_{im}\left(\lambda_\beta\right)_{jn}\left(\lambda_\gamma\right)_{kl}\qquad\text{for $\alpha,\beta,\gamma\in\{0,\dots,8\}$}.
\end{equation*}
Here $H$ is the coupling strength of mass dimension~$-5$, representing the six-fermion vertex interaction.

\end{subsection}

\begin{subsection}{Three-flavor nonlocal NJL model}\label{modelparameter}

Given the four- and six-fermion couplings in the nonlocal framework, we can now write down the Euclidean three-flavor nonlocal NJL action, $\mathcal{S}_\text{E}$, that will be the basis of all calculations performed in this work. We have
\begin{equation}\label{se}
 	\begin{aligned}
 	 	\mathcal{S}_\text{E}&=\int\diff^4 x\left\{\bar\psi(x)\left[-\imu\gamma^\mu\partial_\mu+\hat m_q\right]\psi(x)-\dfrac{G}{2}\left[j_\alpha^S(x)j_\alpha^S(x)+j_\alpha^P(x)j_\alpha^P(x)\right]+\right.\\
		&\qquad\quad\qquad \left. -\dfrac{H}{4}\mathcal{A}_{\alpha\beta\gamma}\left[j_\alpha^S(x)j_\beta^S(x)j_\gamma^S(x)-3j_\alpha^S(x)j_\beta^P(x)j_\gamma^P(x)\right]\right\},
 	\end{aligned}
\end{equation}
where the first term is the kinetic term and $\hat m_q=\text{diag}(m_u,m_d,m_s)$ is the mass matrix with the current quark masses $m_u,m_d,m_s$; $G$ and $H$ are constants to be determined and the densities $j_\alpha^S,j_\alpha^P$ are given in Eq.\,\eqref{j} with the nonlocality distribution $\mathcal{C}(z)$ yet to be specified.

In the remainder of this Section~\ref{model} we demonstrate how this approach works in reproducing zero-temperature QCD, the nonperturbative vacuum and its lowest quark-antiquark excitations: the pseudoscalar meson nonet including decay constants and $\eta$-$\eta'$ mixing. Thermodynamics and the implementation of the Polyakov loop (the step from the nonlocal NJL to the PNJL model) will be described in Section~\ref{thermodynamics} and performed as in Ref.~\cite{Hell} by the gauge covariant replacement $\partial_\mu\to\partial_\mu+\imu A_\mu\delta_{\mu 4}$, and by adding the Polyakov-loop effective potential to the action.

As usual, we start from the partition function
\begin{equation}\label{partitionfunction}
 	\mathcal{Z}=\int\mathscr{D}\bar\psi\,\mathscr{D}\psi\,\euler^{-\mathcal{S}_\text{E}}
\end{equation}
and seek to replace the fermionic fields appearing in Eq.\,\eqref{partitionfunction} by bosonic fields. For this reason, we introduce 18 bosonic fields $\sigma_\alpha$ and $\pi_\alpha$ ($\alpha\in\{0,\dots,8\}$) and, additionally, 18 auxiliary fields $S_\alpha,P_\alpha$ in order to deal with the six-fermion interactions induced by the 't\,Hooft term. 

Inserting a ``one'' in terms of delta functions,
\begin{equation*}
 	1=\int\mathscr{D} S_\alpha\,\mathscr{D} P_\beta\,\delta(S_\alpha-j_\alpha^S)\,\delta(P_\beta-j_\beta^P)=\int\mathscr{D} S_\alpha\,\mathscr{D}P_\beta\,\mathscr{D}\sigma_\alpha\,\mathscr{D}\pi_\beta\,\euler^{\int\diff^4 z\,\sigma_\alpha(S_\alpha-j_\alpha^S)}\,\euler^{\int\diff^4 z\,\pi_\beta(P_\beta-j_\beta^P)},
\end{equation*}
the partition function can be written as
\begin{align*}
 	\mathcal{Z}&=\int\mathscr{D}\bar\psi\,\mathscr{D}\psi\,\mathscr{D} S_\alpha\,\mathscr{D}P_\beta\,\mathscr{D}\sigma_\alpha\,\mathscr{D}\pi_\beta\,\exp\left\{-\int\diff^4x\left[\bar\psi\left(-\imu\gamma^\mu\partial_\mu+\hat m_q\right)\psi+\sigma_\alpha j_\alpha^S+\pi_\beta j_\beta^P\right]\right\}\\
	&\qquad\qquad\times\exp\left\{\int\diff^4 x\left[\frac{G}{2}\left(j_\alpha^Sj_\alpha^S+j_\alpha^P j_\alpha^P\right)+\frac{H}{4}\mathcal{A}_{\alpha\beta\gamma}\left(j_\alpha^Sj_\beta^Sj_\gamma^S-3j_\alpha^S j_\beta^P j_\gamma^P\right)+\sigma_\alpha S_\alpha+\pi_\beta P_\beta\right]\right\}.
\end{align*}
The second term has expressions quadratic and cubic in the densities $j$. Applying the inserted delta functions dictates a replacement of $j_\alpha^S$ and $j_\alpha^P$ by $S_\alpha$ and $P_\alpha$, respectively. The first exponential, on the other hand, contains terms of the form $\bar\psi\hat{\mathscr{A}}\psi$; hence the path integration over the fermionic fields $\bar \psi$ and $\psi$ can be carried out by standard means, leading to
\begin{equation}\label{bosonizedpartitionfunction}
 	\begin{aligned}
 	 	\mathcal{Z}&=\int\mathscr{D}\sigma_\alpha\,\mathscr{D}\pi_\alpha\,\det\hat{\mathscr{A}}\,\int\mathscr{D} S_\alpha\,\mathscr{D} P_\alpha\,\exp\left\{\int\diff^4 x\left(\sigma_\alpha S_\alpha+\pi_\alpha P_\alpha\right)\right\}\\
		&\qquad\qquad\times\exp\left\{\int\diff^4x\left[\dfrac{G}{2}\left(S_\alpha S_\alpha+P_\alpha P_\alpha\right)+\dfrac{H}{4}\mathcal{A}_{\alpha\beta\gamma}\left(S_\alpha S_\beta S_\gamma-3 S_\alpha P_\beta P_\gamma\right)\right]\right\},
 	\end{aligned}
\end{equation}
where $\det\hat{\mathscr{A}}$ is the fermion determinant. In momentum space one finds after a simple Fourier transformation
\begin{equation}\label{fermiondet}
 	\mathscr{A}(p,p'):=\langle p|\hat{\mathscr{A}}| p'\rangle=\left(-\slashed{p}+\hat m_q\right) (2\pi)^4\delta(p-p')+\mathcal{C}\!\left(\frac{p+p'}{2}\!\right)\lambda_\alpha\left[\sigma_\alpha(p-p')+\imu\,\gamma_5\pi_\alpha(p-p')\right].
\end{equation}
Here and in the following we conveniently write $\mathcal{C}(p)\equiv\tilde{\mathcal{C}}(p)$ for the Fourier transform of the distribution $\mathcal{C}(z)$.

\begin{subsubsection}{Stationary phase approximation}
 	Owing to the cubic terms in $S_\alpha$ and $P_\alpha$ the path integration over these fields cannot be carried out explicitly. Therefore, the stationary phase approximation (SPA) is used, choosing the fields $S_\alpha,P_\alpha$ so as to minimize the integrand in the bosonized partition function Eq.\,\eqref{bosonizedpartitionfunction}. Consequently, a necessary condition imposed on the fields is
\begin{equation}\label{spa}
	\begin{aligned}
	 	\sigma_\alpha+G S_\alpha+\frac{3 H}{4}\mathcal{A}_{\alpha\beta\gamma}\left[ S_\beta  S_\gamma- P_\beta  P_\gamma\right]&=0\,,\\
		\pi_\alpha+G P_\alpha-\dfrac{3 H}{2} \mathcal{A}_{\alpha\beta\gamma} S_\beta P_\gamma&=0\,,
	\end{aligned}
\end{equation}
where $S_\alpha, P_\alpha$ are now to be considered as (implicit) functions of $\sigma_\alpha,\pi_\alpha$.
The bosonized action can thus be written as
\begin{equation}\label{bosaction}
	\begin{aligned}
 	\mathcal{S}_\text{E}^\text{bos}&=-\ln\,\det\hat{\mathscr{A}}-\int\diff^4 x\left\{\sigma_\alpha S_\alpha+\pi_\alpha P_\alpha+\dfrac{G}{2}\left[S_\alpha S_\alpha+P_\alpha P_\alpha\right]+\right.\\
	&\qquad\qquad\qquad\qquad\qquad\qquad\qquad\left.+\dfrac{H}{4}\mathcal{A}_{\alpha\beta\gamma}\left[S_\alpha S_\beta S_\gamma-3 S_\alpha P_\beta P_\gamma\right]\right\}.
	\end{aligned}
\end{equation}

\end{subsubsection}

\begin{subsubsection}{Mean-field approximation, gap equations and chiral condensates}
 
As a next step we expand the bosonized action, $\mathcal{S}_\text{E}^\text{bos}$, in a power series around the expectation values of the fields $\sigma_\alpha,\pi_\alpha$,
\begin{equation}
 	\begin{aligned}
 	 	\sigma_\alpha(x)&=\bar\sigma_\alpha+\delta\sigma_\alpha(x),\\
		\pi_\alpha(x)&=\delta\pi_\alpha(x).
 	\end{aligned}
\end{equation}
A first constraint is imposed on the scalar fields by charge conservation; i.\,e., the charge matrix $\hat Q=\text{diag}\left(\frac{2}{3},-\frac{1}{3},-\frac{1}{3}\right)$ commutes with the $\text{SU}(3)$ generators: $[\hat Q,\lambda_\alpha]=0$. This is only possible for $\lambda_0,\lambda_3, \lambda_8$ which means in turn that only $\sigma_0,\sigma_3$ and $\sigma_8$ have to be considered (in the isospin limit which will be investigated later one has the additional constraint that $\sigma_3$ also vanishes). Given these conditions it is useful to introduce
\begin{equation}
 	\sigma=\text{diag}(\sigma_u,\sigma_d,\sigma_s):=\sigma_0\lambda_0+\sigma_3\lambda_3+\sigma_8\lambda_8,
\end{equation}
and, analogously,  $S=\text{diag}(S_u,S_d,S_s)=S_0\lambda_0+S_3\lambda_3+S_8\lambda_8$.

Since we have $\langle\pi_\alpha\rangle=\langle P_\alpha\rangle=0$ to leading order, the action in mean-field approximation reads
\begin{equation}\label{smfa}
 	\dfrac{\mathcal{S}^\text{MF}_\text{E}}{V^{(4)}}=- 2N_\text{c}\int\dfrac{\diff^4 p}{(2\pi)^4}\text{Tr}\,\ln\left[p^2 1_{3\times3}+\hat M^2(p)\right]-\dfrac{1}{2}\Bigg\{\sum_{i\in\{u,d,s\}}\left(\bar\sigma_i \bar S_i+\dfrac{G}{2}\bar S_i\bar S_i\right)+\dfrac{H}{2}\bar S_u\bar S_d\bar S_s\Bigg\},
\end{equation}
where $\hat M(p)=\text{diag}\left(M_u(p),M_d(p),M_s(p)\right)$ with
\begin{equation}
  	M_i(p)=m_i+\bar\sigma_i\mathcal{C}(p),
\end{equation}
 and $1_{3\times3}$ denotes the unity matrix in flavor space and $V^{(4)}$ is the four-dimensional Euclidean volume. 

The mean-field equations (gap equations) are deduced applying the principle of least action, $\frac{\delta\mathcal{S}_\text{E}^\text{MF}}{\delta\sigma_i}=0$ for $\sigma_i=\bar\sigma_i$ ($i\in\{u,d,s\}$). Taking into account that $S_i$ and $P_i$ are both implicit functions of $\sigma_i$, determined through the SPA equations in mean-field approximation (compare Eq.~\eqref{spa}) one obtains the following set of coupled gap equations
\begin{subequations}\label{gaps}
\begin{align}
	\bar\sigma_i&=-G\bar S_i-\dfrac{H}{4}\varepsilon_{ijk}\varepsilon_{ijk}\bar S_{j}\bar S_{k} \label{SPAbasis}\\
		\bar S_i&=- 8 N_\text{c}\int\dfrac{\diff^4 p}{(2\pi)^4}\,\mathcal{C}(p)\,\dfrac{M_i(p)}{p^2+M_i^2(p)}\label{mfeq}.
	\end{align}
\end{subequations}

Finally, the chiral condensate $\langle \bar q q\rangle$ can be calculated from $\mathcal{S}_\text{E}^\text{bos}$ using the Feynman-Hellmann theorem, by differentiation with respect to the current quark mass $m_q$. Equivalently, one can use the definition
\begin{equation}
	\langle\bar q q\rangle=-\imu\,\text{Tr}\,\lim_{y\to x^+}\left[S_\text{F}(x,y)-S_\text{F}^{(0)}(x,y)\right],
\end{equation}
with the full fermion Green function,
\begin{equation}
 	S_\text{F}(x,y)=\int\dfrac{\diff^4 p}{(2\pi)^4}\,\euler^{\imu p(x-y)} S_\text{F}(p)\,,
\end{equation}
and the free quark propagator  $S_\text{F}^{(0)}$ subtracted. Using 
\begin{equation}
	S_{\text{F}}=\dfrac{1}{-\slashed{p}+M_q(p)}
\end{equation}
 this leads to
\begin{equation}\label{chiralcondensate}
\langle \bar q q\rangle=-4  N_\text{c}\int\dfrac{\diff^4 p}{(2\pi)^4}\left[\dfrac{M_q(p)}{p^2+M^2_q(p)}-\dfrac{m_q}{p^2+m_q^2}\right].
	\end{equation}
Note that $M(p)\to m_q$ for large $p$. The subtraction makes sure that no perturbative artifacts are left over in $\langle\bar q q\rangle$ for $m_q\neq0$.

\end{subsubsection}

\begin{subsubsection}{Second-order corrections and meson masses}

In order to calculate the masses of the pseudoscalar mesons, we go beyond mean-field approximation and consider second-order corrections to the mean-field action, extracted from a functional Taylor expansion,
\begin{align*}
 	\mathcal{S}_\text{E}^{(2)}&=\dfrac{1}{2}\int\diff^4 x\,\diff^4 y\,\dfrac{\delta^2\mathcal{S}_\text{E}}{\delta\sigma_\alpha\,\delta\sigma_\beta}\,\delta\sigma_\alpha(x)\,\delta\sigma_\beta(y)+\dfrac{1}{2}\int\diff^4 x\,\diff^4 y\,\dfrac{\delta^2\mathcal{S}_\text{E}}{\delta\pi_\alpha\,\delta\pi_\beta}\,\delta\pi_\alpha(x)\,\delta\pi_\beta(y),
\end{align*}
where the second derivatives, $\frac{\delta^2\mathcal{S}_\text{E}}{\delta\sigma_\alpha(x)\,\delta\sigma_\beta(y)}$ etc., are evaluated at the mean-field values $\sigma_\alpha(x)=\bar\sigma_\alpha$, etc.
We now focus on pseudoscalar mesonic excitations and change the basis according to $\pi_{ij}=\frac{1}{\sqrt{2}}\left(\lambda_\alpha\pi_\alpha\right)_{ij}$. This gives a standard representation of the pseudoscalar meson octet:
\begin{equation}	\pi_{ij}=\left(\hat{\pi}\right)_{ij}=\begin{pmatrix}\frac{\pi^0}{\sqrt{2}}+\frac{\eta_8}{\sqrt{6}}+\frac{\eta_0}{\sqrt{3}}&\pi^+&K^+\\
 	          			\pi^-&-\frac{\pi^0}{\sqrt{2}}+\frac{\eta_8}{\sqrt{6}}+\frac{\eta_0}{\sqrt{3}}&K^0\\
					K^-&\bar K^0&-\frac{2\eta_8}{\sqrt{6}}+\frac{\eta_0}{\sqrt{3}}
 	         \end{pmatrix}.
\end{equation}
Defining analogously a matrix $\hat{\sigma}$ for the scalar mesons, the fermion determinant, Eq.\,\eqref{fermiondet}, can be written as
\begin{equation*}\label{fermdet}
 \hat{\mathscr{A}}(p,p')=\left(-\slashed{p}+\hat m_q\right)(2\pi)^4\delta^{(4)}(p-p')+\mathcal{C}\!\left(\frac{p-p'}{2}\!\right)\sqrt{2}\left[\hat{\sigma}(p-p')+\imu\,\gamma_5\hat{\pi}(p-p')\right].
	\tag{\ref{fermiondet}$'$}
\end{equation*}

Next, we calculate the derivatives appearing in the Taylor expansion. Some caveats of the calculation are outlined in Appendix \ref{secondorderapp}. The resulting second-order contributions to the action are given by
\begin{equation}\label{se2}
 	S_\text{E}^{(2)}=\dfrac{1}{2}\int\dfrac{\diff^4 p}{(2\pi)^4}\left[G^+_{ij,k\ell}(p)\,\delta\sigma_{ij}(p)\,\delta\sigma_{k\ell}(-p)+G^-_{ij,k\ell}(p)\,\delta\pi_{ij}(p)\,\delta\pi_{k\ell}(-p)\right],
\end{equation}
with
\begin{equation}\label{G}
 	G^\pm_{ij,k\ell}(p)=\varPi_{ij}^\pm\,\delta_{i\ell}\,\delta_{jk}+\left(r_{ij,k\ell}^\pm\right)^{-1},
\end{equation}
where
\begin{equation}\label{varPi}
 \varPi_{ij}^\pm(p)=-8 N_\text{c}\int\dfrac{\diff^4 q}{(2\pi)^4}\mathcal{C}^2(q)\dfrac{q^+\cdot q^-\mp M_i(q^+) M_j(q^-)}{\big[{q^+}^2+M_i^2(q^+)\big]\big[{q^-}^2+M_j^2(q^-)\big]},
\end{equation}
$q^\pm=q\pm \frac{p}{2}$, and $(r^\pm)^{-1}$ is defined as the solution of the system
\begin{equation}\label{rsystem}
 	\left[G\delta_{km}\delta_{n\ell}\pm\dfrac{H}{2}\varepsilon_{knt}\varepsilon_{t\ell k n} S_t\right] \left(r_{ij,k\ell}^\pm\right)^{-1}=\delta_{im}\delta_{jn}\ .
\end{equation}

The meson masses can now be determined by writing the second-order term of the action, Eq.~\eqref{se2}, in the physical basis as
\begin{align*}
 	\mathcal{S}_\text{E}^{(2)}\Big|_P&=
\dfrac{1}{2}\int\dfrac{\diff^4 p}{(2\pi)^4}\left\{ G_\pi(p^2)\left[\pi^0(p)\,\pi^0(-p)+2\pi^+(p)\,\pi^-(-p)\right]+\right.\\
		&\qquad+G_K(p^2)\left[2 K^0(p)\,\bar K^0(-p)+2 K^+(p)\,K^-(-p)\right]+\\
		&\qquad+ \left.G_{88}(p^2)\,\eta_8(p)\,\eta_8(-p)+G_{00}(p^2)\,\eta_0(p)\,\eta_0(-p)+2G_{08}(p^2)\,\eta_0(p)\,\eta_8(-p)\right\},
\end{align*}
 where the functions $G_P$ are defined according to Eq.\,\eqref{G}. If one considers only the isospin symmetric case, $m_u=m_d$, then one has
\begin{align*}
 	G_\pi(p^2)&=\left(G+\dfrac{H}{2}\bar S_s\right)^{-1}+\varPi^-_{uu}(p^2)\\
	G_K(p^2)&=\left(G+\dfrac{H}{2}\bar S_u\right)^{-1}+\varPi^-_{us}(p^2)\\
	G_{88}(p^2)&=\dfrac{1}{3}\left[\dfrac{6 G- 4 H \bar S_u-2 H\bar S_s}{2 G^2- G H\bar S_s-H^2 \bar S_u^2}+\varPi^-_{uu}(p^2)+2\varPi^-_{ss}(p^2)\right]\\
	G_{00}(p^2)&=\dfrac{1}{3}\left[\dfrac{6 G+4 H\bar S_u-H\bar S_s}{2 G^2-G H\bar S_s-H^2 \bar S_u^2}+2 \varPi^-_{uu}(p^2)+\varPi^-_{ss}(p^2)\right]\\
	G_{08}(p^2)&=\dfrac{\sqrt{2}}{3}\left[\dfrac{H(\bar S_s-\bar S_u)}{2 G^2- GH\bar S_s- H^2\bar S_u^2}+\varPi^-_{uu}(p^2)-\varPi^-_{ss}(p^2)\right].
\end{align*}
From the construction of the action it is clear that the functions $G_P$ correspond to the inverse pseudoscalar meson propagators. The corresponding masses are given by the poles of these propagators or, equivalently
\begin{equation}\label{mesonmasses}
 	G_P(-m_P^2)=0,\quad\text{for $P\in{\pi,K,\eta}$.}
\end{equation}
Finally, we perform a last basis change in order to produce the physical $\eta$ and $\eta'$ mesons. We introduce the mixing angle $\theta=\theta(p^2)$ and write
\begin{equation}\label{mixingangle}
 	\begin{aligned}
 	 	\eta&=\eta_8\,\cos\theta_\eta-\eta_0\,\sin\theta_\eta\,,\\
		\eta'&=\eta_8\,\sin\theta_{\eta'}+\eta_0\,\cos\theta_{\eta'}\,,
 	\end{aligned}
\end{equation}
where $\theta_\eta=\theta(-m_\eta^2),\theta_{\eta'}=\theta(-m_{\eta'}^2)$.
Introducing the (inverse) $\eta$ and $\eta'$ propagators $G_\eta$ and $G_{\eta'}$, respectively, instead of $G_{00}$,$G_{88}$,$G_{08}$ we obtain for the mixing angle
\begin{equation}
 	\tan2\theta(p^2)=\dfrac{2 G_{08}(p^2)}{G_{00}(p^2)- G_{88}(p^2)}
\end{equation}
and therefore:
\begin{align}
 	G_\eta(p^2)&=\dfrac{G_{88}+G_{00}}{2}-\sqrt{G_{08}^2+\left(\dfrac{G_{00}-G_{88}}{2}\right)^2}\ ,\\
	G_{\eta'}(p^2)&=\dfrac{G_{88}+G_{00}}{2}+\sqrt{G_{08}^2+\left(\dfrac{G_{00}-G_{88}}{2}\right)^2}\ .
\end{align}

\end{subsubsection}

\begin{subsubsection}{Renormalization constants}

One usually introduces renormalized fields\footnote{Here $\varphi(p)$ stands generically for any of the fields $\pi_\alpha(p),\dots$}, $\tilde\varphi(p)=Z_\varphi^{-1/2}\,\varphi(p)$, so that the quadratic part of the Lagrangian can be written as
	\begin{equation*}
	 	\mathscr{L}_\text{E}^{(2)}=\dfrac{1}{2}\left(p^2+m_\phi^2\right)\,\tilde \varphi(p)\,\tilde\varphi(-p),
	\end{equation*}
and it is directly evident that masses are identified with poles of the propagators. From this it is easy to obtain an explicit expression for the renormalization constants, namely
	\begin{equation}\label{Z}
	 	Z^{-1}_P=\left.\dfrac{\diff G_P(p^2)}{\diff p^2}\right|_P,\quad\text{for $P\in{\pi,K,\eta}$.}
	\end{equation}

\end{subsubsection}

\begin{subsubsection}{Decay constants}\label{decayconstantssec}

In this subsection we calculate the (pseudoscalar) meson decay constants defined as
	\begin{equation}\label{fpidef}
	 	\langle0|J_{A,\alpha}^\mu(0)|\tilde\pi_\beta(p)\rangle=\imu\,f_{\alpha\beta}\,p_\mu\quad\Longleftrightarrow\quad\langle0|J_{A,\alpha}^\mu(0)|\pi_\beta(p)\rangle=\imu\,f_{\alpha\beta} Z^{1/2}_\beta\,p_\mu,
	\end{equation}
where $J_{A,\alpha}^\mu(x)=\bar\psi(x)\gamma^\mu\gamma_5\frac{\lambda_\alpha}{2}\psi(x)$ denotes the axial-vector current. In Appendix C of Ref.~\cite{Hell} we outline in some detail the calculation of the (unrenormalized) matrix element $\langle0|J_{A,\alpha}^\mu(0)|\pi_\beta(p)\rangle$.

In order to calculate this matrix element one has to gauge the nonlocal action in Eq.\,\eqref{se}. This requires not only the replacement of the partial derivative by a covariant derivative,
	\begin{equation*}
	 	\partial_\mu\to\partial_\mu+\dfrac{\imu}{2}\gamma_5\lambda_\alpha\,\mathcal{A}_\mu^\alpha(x),
	\end{equation*}
where $\mathcal{A}_\mu^\alpha$  $(\alpha\in\{0\,\dots,8\})$ are a set of axial gauge fields, but also the connection of nonlocal terms through a parallel transport with a Wilson line,
\begin{equation*}
 	\mathcal{W}(x,y)=\mathcal{P}\exp\left\{\dfrac{\imu}{2}\int_0^1\diff\alpha\,\gamma_5\lambda_\alpha\,\mathcal{A}_\alpha^\mu(x+(y-x)\alpha)\,(y_\mu-x_\mu)\right\},
\end{equation*}
where we have chosen a straight line that connects the points $x$ and $y$. This means that expressions of the form $\bar\psi(x)\hat{\mathcal{O}}(z)\psi(y)$ (where $\hat{\mathcal{O}}(z)$ is an arbitrary field operator)  in the action $\mathcal{S}_\text{E}$, Eq.\,\eqref{se}, have to be replaced by $\bar\psi(x)\,\mathcal{W}(x,z)\,\hat{\mathcal{O}}(z)\,\mathcal{W}(z,y)\,\psi(y)$, which guarantees the (local) gauge invariance of the underlying Lagrangian. It turns out that the only term that is eventually affected by the gauging is the fermion determinant $\hat{\mathscr{A}}$, Eq.\,\eqref{fermiondet}, which then becomes, in coordinate space,
\begin{equation}\label{gaugefermiondet} 
		\begin{aligned}
 	\mathscr{A}^\text{G}(x,y)&=\left(-\imu\,\slashed\partial_y+\dfrac{1}{2}\gamma_5\lambda_\alpha\,\slashed{\mathcal{A}}^\alpha+\hat m_q\right)\,\delta(x-y)+\\	&\quad+\mathcal{C}(x-y)\,\mathcal{W}\!\left(x,\frac{x+y}{2}\right)\,\varGamma_\alpha\,\varphi_\alpha\!\left(\frac{x+y}{2}\right)\,\mathcal{W}\!\left(\frac{x+y}{2},y\right).
	\end{aligned}
\end{equation}
Here $\varGamma_\alpha$ stands for either $\varGamma_\alpha=\lambda_\alpha$ or $\varGamma_\alpha=\imu\,\gamma_5\lambda_\alpha$, and $\varphi_\alpha$ accordingly for either a scalar field, $\sigma_\alpha$, or a pseudoscalar field, $\pi_\alpha$.

The desired matrix element then follows from the gauged fermion determinant according to 
	\begin{equation}	\langle0|J_{A,\alpha}^\mu(0)|\pi_\beta(p)\rangle=-\left.\dfrac{\delta^2\,\ln\,\det\mathscr{A}^\text{G}}{\delta\pi_\beta(p)\,\delta\mathcal{A}_\mu^\alpha(t)}\right|_{\mathcal{A}=0\atop t=0}.
	\end{equation}	
After a lengthy evaluation of the functional derivatives, we obtain
\begin{equation}\label{fpiexpr}
 	\begin{aligned} 	\langle0|J_{A,\alpha}^\mu(0)|\pi_\beta(p)\rangle&=2\imu\,\left(\lambda_\alpha^{ij}\lambda_\beta^{ji}+\lambda_\beta^{ij}\lambda_\alpha^{ji}\right)\,\text{tr}\left\{\mathcal{C}(q)\dfrac{q_\mu^+ M_i(q^-)}{\big({q^+}^2+M_j^2(q^+)\big)\big({q^-}^2+M_i^2(q^-)\big)}\right\}+\\	&+	2\imu\,\left(\lambda_\alpha^{ij}\lambda_\beta^{ji}+\lambda_\beta^{ij}\lambda_\alpha^{ji}\right)\,\text{tr}\left\{\int_0^1\diff\alpha\,q_\mu\dfrac{\diff\mathcal{C}(q)}{\diff q^2}\dfrac{M_i(q_\alpha^+)}{{q_\alpha^+}^2+M_i^2(q_\alpha^+)}\right\} 	
 	+\\
	&+2\imu\,\bar\sigma_j\left(\lambda_\alpha^{ij}\lambda_\beta^{ji}+\lambda_\beta^{ij}\lambda_\alpha^{ji}\right)\times\\
		&\qquad\times\text{tr}\left\{\int_0^1\diff\alpha \,q_\mu\dfrac{\diff\mathcal{C}(q)}{\diff q^2}\,\mathcal{C}\!\left(\!q\!-\!\frac{p}{2}\alpha\!\right)\dfrac{q_\alpha^+\cdot q_\alpha^-\!+\!M_j(q_\alpha^+)M_i(q_\alpha^-)}{\big({q_\alpha^+}^2\!+\!M_j^2(q_\alpha^+)\big)\big({q_\alpha^-}^2\!+\!M_i^2(q_\alpha^-)\big)}\right\},
 	\end{aligned}
\end{equation}
with
\begin{equation}\label{palpha}
 	\begin{aligned}
 	 	q_\alpha^+&=q+\frac{p}{2}(1-\alpha),\qquad& q_\alpha^-&=q-\frac{p}{2}(1+\alpha)\\
		q^+&=q+\frac{p}{2},& q^-&=q-\frac{p}{2}\,.
 	\end{aligned}
\end{equation}
Now, the decay constants can be derived from the expression \eqref{fpiexpr} and their definitions, Eq.\,\eqref{fpidef}, by contraction with $p^\mu$, hence
\begin{equation}\label{decayconstants}
 	f_{\alpha\beta}=\imu\,p_\mu\langle0|J_{A,\alpha}^\mu(0)|\pi_\beta(p)\rangle \dfrac{Z_\beta^{-1/2}}{m_\beta^2},
\end{equation}
evaluated at the corresponding mass $p^2=-m_\beta^2$.
Owing to the properties of the Gell-Mann matrices one has $f_{\alpha\beta}=\delta_{\alpha\beta} f_\pi$ for $\alpha\in\{1,2,3\}$ and $f_{\alpha\beta}=\delta_{\alpha\beta} f_K$ for $\alpha\in\{4,5,6,7\}$. On the other hand, for the $0$- and $8$ component we obtain
\begin{align*}
 	f_{88}(p^2)&=\dfrac{4}{3}\left[2 f_{ss}(p^2)+f_{uu}(p^2)\right]\\
	f_{00}(p^2)&=\dfrac{4}{3}\left[2 f_{uu}(p^2)+f_{ss}(p^2)\right]\\
	f_{08}(p^2)=f_{80}(p^2)&=\dfrac{4\sqrt{2}}{3}\left[f_{uu}(p^2)-f_{ss}(p^2)\right].
\end{align*}

\end{subsubsection}

\end{subsection}

\begin{subsection}{Chiral low-energy theorems}

In this section we derive the Goldberger-Treiman and Gell-Mann--Oakes--Renner relations explicitly from the nonlocal NJL model presented in this work. For this aim, we expand the meson self-energy contribution $\varPi_{uu}^-$, Eq.~\eqref{varPi}, up to first order in the current quark mass $m_u$ and the momentum $p^2$:
\begin{equation}\label{varPiexp}
\varPi_{uu}^-(p^2,m_u)=\dfrac{\bar S_{u,0}}{\bar \sigma_{u,0}}-\dfrac{2\langle \bar uu\rangle_0}{\bar\sigma_{u,0}^2} m_u+Z^{-1}_{\pi,0}\, p^2.
\end{equation}
The first term on the right-hand side follows immediately from Eq.~\eqref{varPi} by setting $m_u=p^2=0$ and using the gap equation \eqref{mfeq} in the chiral limit, $m_u=0$ (index ``$0$"). The second term can be recovered by writing $\mathcal{C}(p)=\frac{1}{\bar\sigma_u}\left(M_u(p)-m_u\right)$ in Eq.~\eqref{varPi} and using the definition of the chiral condensate, Eq.~\eqref{chiralcondensate}. The last term follows from the definition of the renormalization constant $Z_\pi$, Eq.~\eqref{Z}.

Next we notice by expanding the pion decay constant, Eq.~\eqref{fpiexpr}, that only the term in the first line of this equation contributes to order $\mathcal{O}(p^2)$. One finds
\begin{equation}
	\lim_{p^2\to0} f_{uu}(p^2)=\dfrac{1}{4}\bar\sigma_{u,0} Z_{\pi,0}^{-1}\,.
\end{equation}
Using Eq.~\eqref{decayconstants} this implies
\begin{equation}
	f_{\pi,0}=\bar\sigma_{u,0} Z^{-1}_{\pi,0}\,,
\end{equation}
which is nothing but the Goldberger-Treiman relation.
Finally, multiplying both sides of the expansion \eqref{varPiexp} with $G+\frac{H}{2}$ and identifying the pion-mass definition, Eq.~\eqref{mesonmasses}, on the left-hand side, and the gap equation \eqref{SPAbasis}, on the right-hand side, we get
\begin{equation}\label{gmor}
	f_{\pi,0}^2 m_\pi^2=-m_u\langle \bar uu+\bar dd\rangle_0,
\end{equation}
which is the Gell-Mann--Oakes--Renner relation.

We have thus demonstrated that our nonlocal NJL model is consistent, as expected, with fundamental low-energy theorems based on chiral symmetry. At the same time the nonlocal model reduces to the local NJL model results when $\mathcal{C}(p)$ is chosen as a theta function.

\end{subsection}

\begin{subsection}{Parameter fixing and numerical results}\label{parameters}

Now that all relevant formulas have been derived in the preceding subsections, we fix the model parameters in order to quantitatively reproduce physical observables. Apart from the coupling strengths $G$ and $H$ of the four- and six-fermion interaction vertices, this involves a specification of the nonlocality distribution $\mathcal{C}$. In preparation of this input we are guided by results of Schwinger-Dyson QCD calculations and return briefly to the issue of the separable ansatz for $\mathcal{C}$ used in the present approach.

\begin{subsubsection}{Schwinger-Dyson calculations vs separable approximation}\label{sdsepcomparison}

For simplicity we restrict ourselves in this subsection to the case of two degenerate (up- and down-quark) flavors and $H=0$. The full SD equation is then given by Eq.~\eqref{sde},  while the separable equation follows by replacing $\tilde{\mathcal{C}}(q-p)\to\mathcal{C}(q)\cdot\mathcal{C}(p)$. In this case, the mass equation can be written as $M(q)=m_u+\bar\sigma_u\mathcal{C}(q)$ (see Eqs.~\eqref{smfa} and \eqref{SPAbasis}), with $m_u=m_d$ the current quark masses and $\bar\sigma_u=\bar\sigma_d$ the mean-field values of the $\sigma$ fields. Furthermore, for this test case and exclusively in this subsection, we choose a Gaussian for the momentum distribution function $\mathcal{C}$ in both the full SD expression and the separable ansatz. The parameters $G$, $m_u$ and the width of the Gaussian are fixed such as to reproduce the pion mass, $m_\pi=140\,\text{MeV}$,  the pion decay constant\footnote{For the SD equation we use the formulae given in Ref.~\cite{Praschifka}, while for the separable ansatz we use formulae \eqref{mesonmasses} and \eqref{decayconstants}.}, $f_\pi=92\,\text{MeV}$ and the chiral condensates ranging between $\langle\bar uu\rangle=\langle\bar dd\rangle\approx -(260\,\text{MeV})^3$ and $-(280\,\text{MeV})^3$. 

Figure~\ref{DSEplot} shows results from the full SD and separable nonlocal NJL approaches in comparison. It is evident that the results coincide at the five-percent level. One should recall at this point that the volume element in four dimensions is $\diff^4p\sim p^3\,\diff p$, so that details of the low momentum behavior of $\mathcal{C}(p)$ in the integrand do not dramatically influence the value of the integral in the gap equation.

\begin{figure}
\begin{center}
\begin{minipage}[t]{.475\textwidth}{
\includegraphics[width=\textwidth]{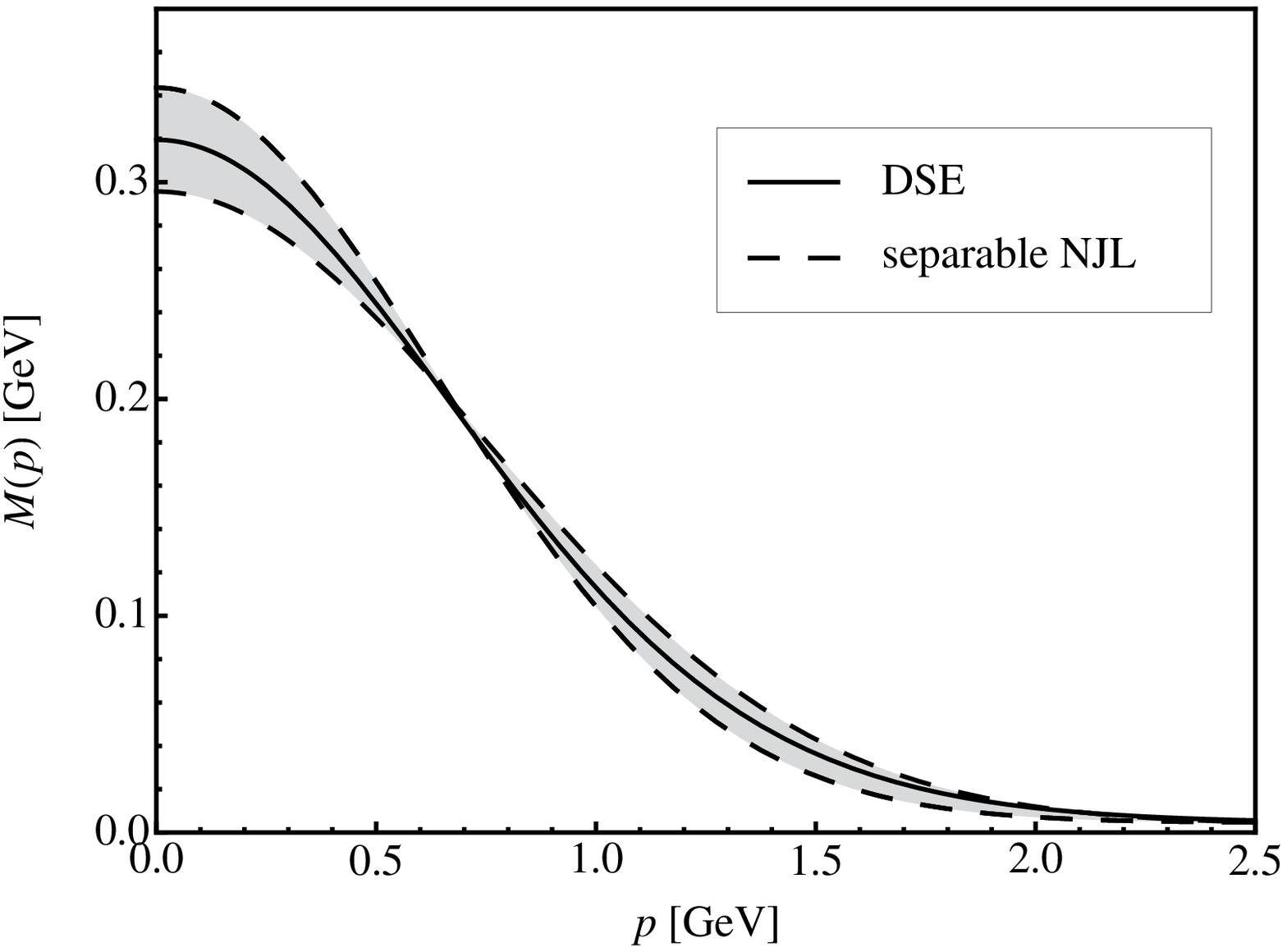}
\caption{Comparison of the self-consistent solution of the Schwinger-Dyson equation \eqref{sde} (solid line) to the solution of the gap equations \eqref{gaps} derived for a separable interaction (dashed lines) for values of the chiral condensate $\langle\bar uu\rangle$ ranging between $-(260\,\text{MeV})^3$ and $-(280\,\text{MeV})^3$ (gray band).}\label{DSEplot}
}\end{minipage}\hfill
\begin{minipage}[t]{.475\textwidth}{
 \includegraphics[width=\textwidth]{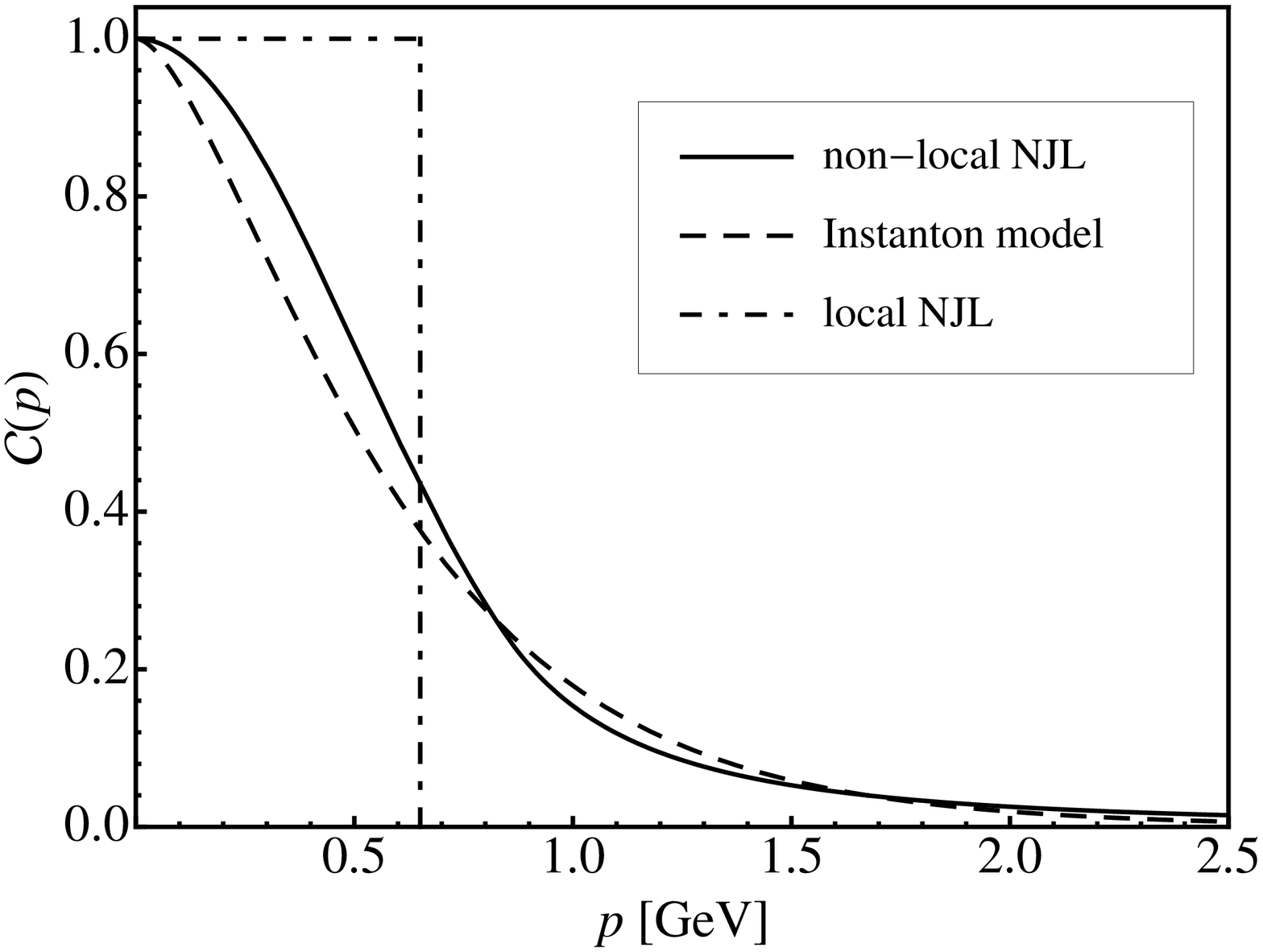}
		\caption{Comparison of the distribution $\mathcal{C}(p)$ used in the nonlocal NJL model (solid line) and the one derived from the instanton model with instanton size $d=0.35\,\text{fm}$ (dashed line). The dash-dotted line shows the function $\mathcal{C}(p)$ in a local NJL model with Euclidean four-momentum cutoff.\label{CvsCI}}}
\end{minipage}
\end{center}
\end{figure}

\end{subsubsection}

\begin{subsubsection}{Parameters of the nonlocal 3-flavor NJL model}

Apart from the nonlocality distribution $\mathcal{C}(p)$ with its characteristic scale, the parameters to be fixed are the coupling strengths $G$ and $H$ and the current quark masses $m_u$ ($=m_d$) and $m_s$. Consider first the distribution $\mathcal{C}(p)$ again. Its asymptotic form is constrained by QCD through the high-momentum behavior of the quark mass function $M(p)$ to leading order in the operator product expansion:
\begin{equation*}
 	M(p)\propto-\dfrac{\alpha_\text{s}(p^2)}{p^2}\,\langle\bar u u+\bar dd\rangle\quad\text{at $p>\varLambda\sim1\,\text{GeV}$.}
\end{equation*}
We use the leading order expression $\alpha_\text{s}(p^2)= 4\pi \Big[\beta_0\ln\frac{p^2}{\Lambda^2_\text{QCD}}\Big]^{-1}$ which is sufficient for our purposes. We set $\beta_0=9$ (with three active flavors) and $\Lambda_\text{QCD}=0.25\,\text{GeV}$ which reproduces $\alpha_\text{s}=0.12$ at $m_Z=91.2\,\text{GeV}$. At smaller momentum scales we are guided by results from Schwinger-Dyson calculations (in Landau gauge) and by extrapolations of lattice QCD data for $M(p)$ \cite{Bowman}.

An alternative choice is to use an instanton model. All these options lead to very similar shapes of $\mathcal{C}(p)$. For the actual calculations we adopt the parametrization introduced in Ref.~\cite{Hell}:
\begin{equation}\label{c}
 	\mathcal{C}(p^2)=\begin{cases} \euler^{-p^2 d^2/2}&\text{for $p^2<\varLambda^2$}\\
 	                  		\text{const.}\cdot\dfrac{\alpha_\text{s}(p^2)}{p^2}&\text{for $p^2\ge\varLambda^2$}\,.
 	                 \end{cases}
\end{equation}
The normalization condition $\mathcal{C}(p=0)=1$ and the matching at $p=\varLambda$ fix the remaining constant once $d$ is chosen. Setting $d=0.35\,\text{fm}$ guided by a typical instanton size, $\mathcal{C}(p)$ closely resembles the instanton based distribution \eqref{instantonc}, as demonstrated in Fig.~\ref{CvsCI}. The resulting matching scale in Eq.~\eqref{c} is $\varLambda=0.85\,\text{GeV}$.

Having determined $\mathcal{C}(p)$ we proceed by reproducing masses and decay constants of the pseudoscalar meson nonet. Two scenarios with marginally different coupling strengths $G$ (leaving $\mathcal{C}(p)$ and all remaining parameters unchanged) will be considered. ``Scenario~I'' optimizes the $\eta'$ sector including $\eta$-$\eta'$ mixing. ``Scenario~II'' provides a best fit to pseudoscalar octet observables.

\begin{table}
\begin{center}
\begin{tabular}{|c|c|c|c|}\hline\hline
 $G$&$H$&$m_u$&$m_s$\\\hline\hline
$(0.96\,\text{fm})^2$&$-(0.63\,\text{fm})^5$&$3.0\,\text{MeV}$&$70\,\text{MeV}$\\\hline
\end{tabular}
\caption{Scenario~I parameter set of the $N_\text{f}=3$ nonlocal NJL model.}
\label{paramsI}
\end{center}
\end{table}

\begin{table}
\begin{center}
\begin{tabular}{|c|c|c|c|}\hline\hline
 $\langle\bar uu\rangle=\langle\bar dd\rangle$&$\langle\bar ss\rangle$&$M_u=M_d$&$M_s$\\\hline\hline
$-(0.282\,\text{GeV})^{3}$&$-(0.303\,\text{GeV})^3$&$362\,\text{MeV}$&$575\,\text{MeV}$\\\hline\hline
\end{tabular}
\\
\vspace{.3cm}
\begin{tabular}{|c|c|c|c|c|c|c|c|}\hline\hline
 $m_\pi$&$m_K$&$m_\eta$&$m_{\eta'}$&$f_\pi$&$f_K$&$\theta_\eta$&$\theta_{\eta'}$\\\hline\hline
$138\,\text{MeV}$&$487\,\text{MeV}$&$537\,\text{MeV}$&$954\,\text{MeV}$&$83.4\,\text{MeV}$&$104.1\,\text{MeV}$&$3.3^\circ$ &$-29.1^\circ$\\\hline\hline
\end{tabular}
\caption{Calculated physical quantities using the scenario~I parameters of Table~\ref{paramsI}.}
\label{resultsI}
\end{center}
\end{table}

Choosing the parameters of scenario~I as given in Table~\ref{paramsI}, one finds the values of the pseudoscalar masses\footnote{Note, that the $\bar uu$-threshold is lower than the $\eta'$ mass. Hence, the integrals determining the $\eta'$-mass might be ill-defined owing to poles in the integration region. Therefore, in fixing the $\eta'$-mass, we apply the regularization method described in Refs.~\cite{Cutkosky,Scarpettini}.}, decay constants and  $\eta$-$\eta'$-mixing angle as shown in Table~\ref{resultsI}. The current quark masses are consistent with those listed in the PDG table \cite{PDG} at a renormalization scale of about $2\,\text{GeV}$. The $\eta'$ mass is very close to its experimental value. The same is true for the ratio of the decay constants $f_K/f_\pi=1.25$ (compared to the experimental $(f_K/f_\pi)_\text{exp}=1.22$). The pion decay constant $f_\pi$, though, is approximately 10\,\% off its experimental value but close to its value at the chiral limit. 

Furthermore, it is instructive to compare our result for the $\eta$-$\eta'$-mixing angle, $\theta_{\eta'}=-29.1^\circ$, to the empirical value. The most recent analysis \cite{DiMicco} gives\footnote{Note the different definitions of the $\eta$-$\eta'$-mixing angle in this work and in Ref.~\cite{DiMicco}. The cited number $\theta=-29.0^\circ$ has, however, already been translated to the definition, Eq.~\eqref{mixingangle}, of the mixing angle used in the present work.} $\theta=-29.0^\circ$ and agrees perfectly with our result. Note, however, that in Ref.~\cite{DiMicco} contributions  from the gluon condensate are included which our model does not explicitly account for.
The left part of Fig.~\ref{sigmarunning}  shows the momentum dependence of the resulting dynamical up-quark mass, $M_u(p)$, compared to lattice data from Ref.~\cite{Bowman}.

\begin{table}
\begin{center}
\begin{tabular}{|c|c|c|c|}\hline\hline
 $G$&$H$&$m_u$&$m_s$\\\hline\hline
$(1.04\,\text{fm})^2$&$-(0.63\,\text{fm})^5$&$3.0\,\text{MeV}$&$70\,\text{MeV}$\\\hline
\end{tabular}
\caption{Scenario~II parameter set of $N_\text{f}=3$ nonlocal NJL model.}
\label{paramsII}
\end{center}
\end{table}

\begin{table}
\begin{center}
\begin{tabular}{|c|c|c|c|}\hline\hline
 $\langle\bar uu\rangle=\langle\bar dd\rangle$&$\langle\bar ss\rangle$&$M_u=M_d$&$M_s$\\\hline\hline
$-(0.304\,\text{GeV})^{3}$&$-(0.323\,\text{GeV})^3$&$468\,\text{MeV}$&$694\,\text{MeV}$\\\hline\hline
\end{tabular}
\\
\vspace{.3cm}
\begin{tabular}{|c|c|c|c|c|c|c|c|}\hline\hline
 $m_\pi$&$m_K$&$m_\eta$&$m_{\eta'}$&$f_\pi$&$f_K$&$\theta_\eta$&$\theta_{\eta'}$\\\hline\hline
$139\,\text{MeV}$&$495\,\text{MeV}$&$547\,\text{MeV}$&$964\,\text{MeV}$&$92.8\,\text{MeV}$&$110.1\,\text{MeV}$&$1.9^\circ$ &$-22.3^\circ$\\\hline\hline
\end{tabular}
\caption{Calculated physical quantities using the parameters of Table~\ref{paramsII}. (These values together with the parameters of Table~\ref{paramsII} are referred to as ``scenario~II''.)}
\label{resultsII}
\end{center}
\end{table}

The parameters of scenario~II (Table~\ref{paramsII}) differ from those of scenario~I only by a four-fermion coupling constant $G$ that is about 15~\% larger.
We see from the calculated quantities given in Table \ref{resultsII} that the pseudoscalar meson masses and decay constants now agree perfectly with their empirical values. On the other hand, the magnitudes of the chiral condensates and the dynamical quark masses $M_u(0)$ and $M_s(0)$ increase, making them less compatible with common phenomenology. This can  easily be understood from the Gell-Mann--Oakes--Renner relation \eqref{gmor} recalling that the current quark and pion masses for scenario~I are the same as for  scenario~II, while the value of the pion decay constant of scenario~II is increased. 
The momentum dependence of the dynamical quark mass, $M(p)$, is shown on the right-hand side of Fig.~\ref{sigmarunning}. In particular, the $\eta$-$\eta'$-mixing angle $\theta_{\eta'}=-22.3^\circ$ now differs by 20\,\% from the deduced empirical value in Ref.\,\cite{DiMicco}.

Finally, a comparison with the standard local NJL model is instructive. From Refs.~\cite{Vogl,Klimt1,Klimt2,Rehberg} one finds the gap equations
\begin{align*}
	M_u&=m_u-\tilde{G}\langle \bar uu\rangle-\dfrac{\tilde{H}}{2}\langle \bar uu\rangle\langle\bar ss\rangle\\
	M_s&=m_s-\tilde{G}\langle \bar ss\rangle-\dfrac{\tilde{H}}{2}\langle \bar uu\rangle^2.
\end{align*}
The equivalent coupling strengths $\tilde{G}$ and $\tilde{H}$ of the local model can be evaluated by comparison with Eq.~\eqref{SPAbasis}. One derives for scenario~I: $\tilde{G}=G\frac{\bar S_u}{\langle\bar uu\rangle}\approx 11\,\text{GeV}^{-2}$ and $\tilde{H}=H\frac{\bar S_u\bar S_s}{\langle \bar uu\rangle\langle\bar ss\rangle}\approx 400\,\text{GeV}^{-5}$, values that lie well in the ballpark of typical local approaches \cite{Vogl,Ratti1,Simon1,Ratti2,Simon2,Klimt1,Klimt2,Rehberg}.

We will comment further on the influence of the different model parameters in Sec.~\ref{comparison} when dealing with thermodynamics resulting from the nonlocal approach.

\begin{figure}[t]
\begin{center}
	\begin{minipage}[t]{.475\textwidth}{\includegraphics[width=\textwidth]{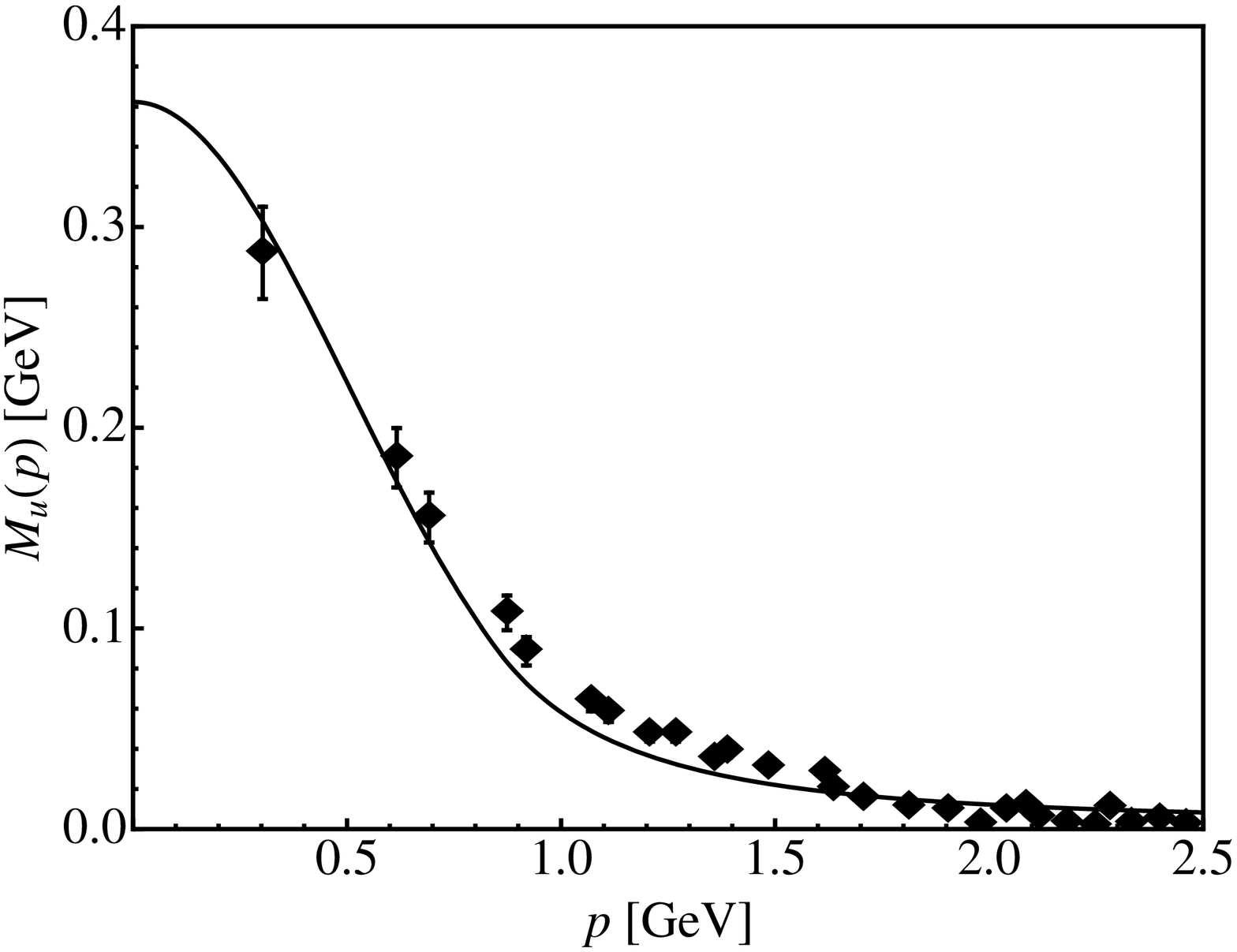}}	
	\end{minipage}
\hfill
	\begin{minipage}[t]{.475\textwidth}{
		\includegraphics[width=\textwidth]{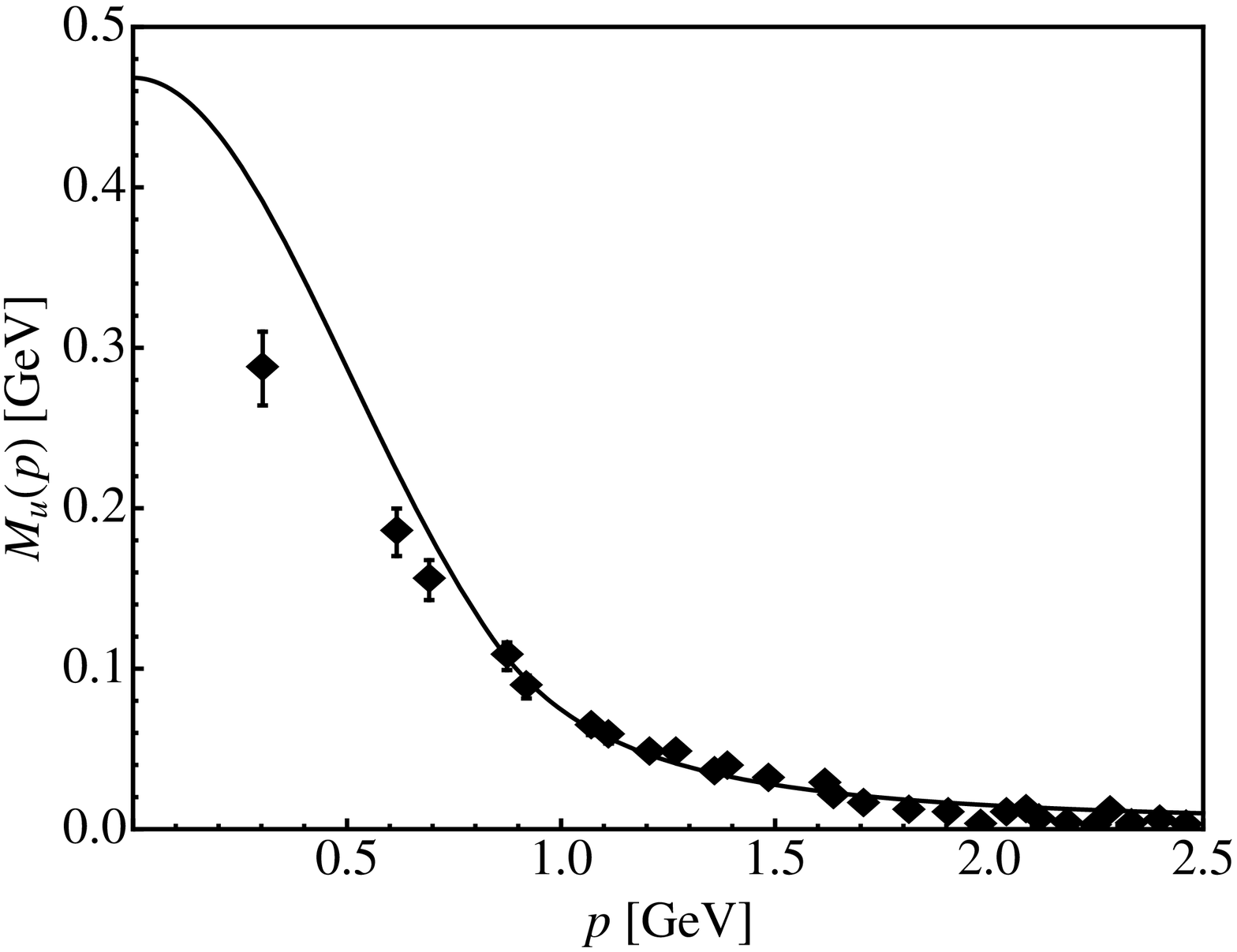}
		}
	\end{minipage}
\end{center}
\vspace{-.75cm}
\caption{Momentum dependence of the dynamical (constituent) up-quark mass  $M_u$ compared with lattice data extrapolated to the chiral limit (from Ref.~\cite{Bowman}). The left figure shows the result for scenario~I, the right figure the result for scenario~II.\label{sigmarunning}}
\end{figure}

\end{subsubsection}

\end{subsection}

\end{section}

\begin{section}{Thermodynamics of the nonlocal PNJL model}\label{thermodynamics}

In this section we apply the nonlocal NJL scheme in the modeling of QCD thermodynamics. Temperature $T$ and quark chemical potential $\mu$ are introduced using the Matsubara formalism \cite{Kapusta}. Important features of the confinement-deconfinement transition are introduced by the Polyakov loop which serves as an  order parameter for  confinement in the pure gauge case \cite{McLerran,Weiss}. Coupling the NJL quarks to a gluonic background field, expressed in terms of the Polyakov loop, allows describing both the chiral and the deconfinement transitions simultaneously. The framework for this is the Polyakov-loop-extended nonlocal Nambu--Jona-Lasinio (nonlocal PNJL) model.

\begin{subsection}{Polyakov loop and its effective potential}

The renormalized Polyakov loop, $\langle\Phi\rangle$, is defined through
\begin{equation}
	\langle\Phi(\vec{x}\,)\rangle=\dfrac{1}{N_\text{c}}\langle\text{tr}_\text{c}\left[L(\vec{x}\,)\right]\rangle\,,
\end{equation}
where $\text{tr}_\text{c}$ denotes the trace over color only and $L$ is the  Polyakov loop, a timelike Wilson line connecting two points at $t=0$ and $t=-\imu\beta$ in imaginary time, with periodic boundary conditions:
\begin{equation}
	L(\vec{x}\,)=\mathcal{P}\exp\left\{\imu\int_0^\beta\diff\tau\, A_4(\tau,\vec{x}\,)\right\}\ .
\end{equation}
Here $\mathcal{P}$ is the path-ordering operator and $A_4:=A_4^a t_a=\imu A_0^a t_a$ is the fourth component (in Euclidean space) of the gluon field\footnote{The color $\text{SU}(3)$ matrices $t_a,a\in\{1,\dots,8\}$ are defined through the Gell-Mann matrices $\lambda_a,a\in\{1,\dots,8\}$ as $t_a:=\frac{\lambda_a}{2}$.}, and $\beta=1/T$.

As shown e.\,g. in Ref.~\cite{McLerran},  $\langle\Phi\rangle=0$ implies an infinite free energy,  corresponding to confinement, while $\langle\Phi\rangle=1$ implies a vanishing free energy of the  system and, thus, deconfinement. According to  Ref.~\cite{Weiss}, $\langle\Phi\rangle$ strictly serves as an  order parameter for the confinement-deconfinement phase transition only in absence of quarks. But even when quarks are present $\langle\Phi\rangle$ is still useful as an indicator for a rapid crossover transition.

Without loss of generality we can restrict ourselves to the diagonal elements of the $\text{SU}(3)$ Lie algebra (the so-called Polyakov gauge). We neglect spatial fluctuations and treat $A_4^3$ and $A_4^8$ as constant fields, so that the Wilson line \eqref{wilson} simplifies to a phase factor $\euler^{\imu (\vec{x}-\vec{y})}\cdot\euler^{\imu (x_4-y_4)(A_4^3 t_3+A_4^8 t_8)}$. 
The contribution of the nondiagonal elements of the $\text{SU}(3)$ Lie algebra can be included by carrying out the group integration, leading to the following Haar volume
\begin{equation}\label{JHaar}
	J(\phi_3,\phi_8)=\dfrac{1}{V_{\text{SU}(3)}}\int\prod_{i\in\{1,2,4,5,6,7\}}\diff \phi_i=\dfrac{2}{3\pi^2}\left(\cos(\phi_3)-\cos(\sqrt{3}\phi_8)\right)^2\sin^2(\phi_3)\,,
\end{equation}
where we have set $\phi_{3,8}=\beta \frac{A_4^{3,8}}{2}$.
This volume can be written in terms  of the Polyakov loop $\Phi$ and its conjugate $\Phi^*$,
\begin{equation}
 	J(\Phi,\Phi^*)=\dfrac{9}{8\pi^2}\left[1-6\Phi^*\Phi+4\left({\Phi^*}^3+\Phi^3\right)-3\left(\Phi^*\Phi\right)^2\right],
\end{equation}
with $\Phi=\frac{1}{N_\text{c}}\text{tr}_\text{c}\left[\exp\left(\imu(\phi_3\lambda_3+\phi_8\lambda_8)\right)\right]$.  This procedure leads to  the construction of the following effective potential $\mathcal{U}$ \cite{Fukushima1,Fukushima2} that incorporates the effects of the six nondiagonal gluon fields:
\begin{equation}\label{polyakovU}
	\dfrac{\mathcal{U}(\Phi,\Phi^*,T)}{T^4}=-\dfrac{1}{2} b_2(T)\Phi^*\Phi+b_4(T)\ln\left[1-6\Phi^*\Phi+4\left({\Phi^*}^3+\Phi^3\right)-3(\Phi^*\Phi)^2\right]\,.
\end{equation}
The coefficients are parametrized as
\begin{equation*}\begin{aligned}b_2(T)&=a_0+a_1\left(\dfrac{T_0}{T}\right)+a_2\left(\dfrac{T_0}{T}\right)^2+a_3\left(\dfrac{T_0}{T}\right)^3\\
b_4(T)&=b_4\left(\dfrac{T_0}{T}\right)^3.
\end{aligned}
\end{equation*}
The first term on the right-hand side is reminiscent of a Ginzburg-Landau ansatz. The values of the coefficients are taken from Ref.~\cite{Simon1} and listed  in Table~\ref{polyakovparameters}. The potential $\mathcal{U}$ is manifestly invariant under transformations with elements of $Z(3)$, the center of $\text{SU}(3)$, which is the underlying symmetry behind the confinement-deconfinement phase transition.
 
The parametrization \eqref{polyakovU} of the Polyakov-loop effective potential $\mathcal{U}$ is applicable at temperatures $T$ up to about twice the critical $T_c$. At higher temperatures, transverse gluon degrees of freedom -- not covered by the Polyakov loop -- begin to be important. Other parametrizations of $\mathcal{U}$ are possible. An example is the two-parameter ansatz \cite{Fukushima3} based on the strong coupling limit. These two versions of $\mathcal{U}$ differ at high temperatures but produce very similar pressure profiles \cite{Fukushima3} at $T\lesssim 2 T_c$, the temperature region of primary interest in the present study.
\begin{table}
\begin{center}
\begin{tabular}{|c|c|c|c|c|c|}
\hline\hline
$a_0$&$a_1$&$a_2$&$a_3$&$b_4$\\\hline\hline
$3.51$&$-2.56$&$15.2$&$-0.62$&$-1.68$\\\hline\hline
\end{tabular}
\caption{Parameters of the Polyakov potential $\mathcal{U}$ (from Ref.~\cite{Simon1}).}\label{polyakovparameters}
\end{center}
\end{table}

\end{subsection}

\begin{subsection}{Coupling of quarks and Polyakov loop}

The coupling of the quarks and the Polyakov loop\footnote{From here onward we omit angled brackets for notational simplicity, i.\,e.~we write $\langle\Phi\rangle\to\Phi$.} $\Phi$ is introduced by the minimal gauge coupling procedure applied to nonlocal field theories. This implies, first, a replacement of the partial derivative $\partial_\mu$ by a covariant derivative $D_\mu=\partial_\mu-\imu A_\mu$, or 
in momentum space $p_\mu\to p_\mu+A_\mu$.\footnote{The coupling strength $g$ is absorbed in the definition of the fields $A_\mu$.} 
Furthermore, in close analogy to the gauging of theories on a discrete lattice (see e.\,g.\ Ref.~\cite{Rothe}), one introduces the gluonic fields writing a Wilson line (compare Sec.~\ref{decayconstantssec})
\begin{equation}\label{wilson}
	\mathcal{W}_A(x,y)=\mathcal{P}\left\{\exp\left[\imu\int_x^y\diff s_\mu t_a A_\mu^a\right]\right\}
\end{equation}
between the (nonlocal) fermionic bilinears, i.\,e.\ $\bar\psi(x)\psi(y)\to\bar\psi(x)\mathcal{W}_A(x,y)\psi(y)$.

In the present context we set
\begin{equation*}
	A_\mu=\delta_{\mu4}(A_4^3 t_3+A_4^8 t_8)\,,
\end{equation*}
following our previous discussion, with constant fields $A_4^{3,8}$. The gauge invariant replacement in quark momentum space is then simply
\begin{equation}\label{gauging}
	p_4\to p_4-(A_4^3 t_3+A_4^8 t_8)\,
\end{equation}
keeping the three-momentum $\vec{p}$ unchanged.

The next step is now the application of the Matsubara formalism as described in the standard literature (e.\,g., Ref.~\cite{Kapusta}). The variable $p_4$ is replaced by the fermionic Matsubara frequency $\omega_n=(2n+1)\pi T,n\in\Z$ and taking into account the gauging \eqref{gauging}. The resulting thermodynamic potential in mean-field approximation is\footnote{Note an extra factor $\frac{1}{2}$ because of the doubling of the degrees of freedom in Nambu-Gor'kov space.}
\begin{equation}\label{OmegaMatsubara}
\begin{aligned}
	\Omega&=-\dfrac{T}{2}\sum_{n\in\Z}\int\dfrac{\diff^3 p}{(2\pi)^3}\,\text{tr}\,\ln\left[\beta\tilde S^{-1}(\imu\omega_n,\vec{p}\,)\right]\\
	&\quad-\dfrac{1}{2}\Bigg\{\sum_{f\in\{u,d,s\}}\left(\bar\sigma_f \bar S_f+\dfrac{G}{2}\bar S_f\bar S_f\right)+\dfrac{H}{2}\bar S_u\bar S_d\bar S_s\Bigg\}+\mathcal{U}(\Phi,\Phi^*,T)\,,
	\end{aligned}
\end{equation}
with
\begin{equation}\label{nambugorkovS}
	\tilde S^{-1}(\imu\omega_n,\vec{p}\,)=\begin{pmatrix}\imu\omega_n\gamma_0-\vec{\gamma}\cdot\vec{p}-\hat M-\imu (A_4+\imu \hat\mu)\gamma_0&0\\ 0&\imu\omega_n\gamma_0-\vec{\gamma}\cdot\vec{p}-\hat M^*+\imu (A_4+\imu \hat\mu)\gamma_0
		                               \end{pmatrix}
\end{equation}
where the momentum-dependent dynamical mass matrix $\hat M$ is diagonal in color and flavor space, 
\begin{equation*}
\hat M=
	\begin{pmatrix}
	\hspace{-15em}\text{diag}_\text{c}(M(\omega_{u,n}^-,\vec{p}\,),M(\omega_{u,n}^+,\vec{p}\,),M(\omega_{u,n}^0,\vec{p}\,))\\[.5ex]
      \text{diag}_\text{c}(M(\omega_{d,n}^-,\vec{p}\,),M(\omega_{d,n}^+,\vec{p}\,),M(\omega_{d,n}^0,\vec{p}\,))\\[.5ex]
	\hspace{15em}\text{diag}_\text{c}(M(\omega_{s,n}^-,\vec{p}\,),M(\omega_{s,n}^+,\vec{p}\,),M(\omega_{s,n}^0,\vec{p}\,))
       \end{pmatrix}
\end{equation*}
 with $\omega_{f,n}^\pm=\omega_n-\imu\mu_f\pm A_4^3/2-A_4^8/(2\sqrt{3})$, $\omega_{f,n}^0=\omega_n-\imu\mu_f+A_4^8/\sqrt{3}$. Note, that a quark chemical potential $\hat\mu=\text{diag}_\text{f}(\mu_u,\mu_d,\mu_s)$ has been introduced that will, however, become important in Sect.~\ref{finitedensity}.\footnote{The chemical potential enters the 4-component of the argument of the mass function according to the same reasoning applied above for the gauging of the model, taking into account that $\mu$ can be considered as an imaginary potential in Euclidean space.}   The trace may be further simplified leading to
\begin{equation*}\label{omegasimpel}
\begin{aligned}
	\Omega&=-2 T\sum_{f\in\{u,d,s\}}\sum_{i=0,\pm} \sum_{n\in\Z}\int\dfrac{\diff^3 p}{(2\pi)^3}\,\text{Re}\left\{\ln\left[{\omega_{f,n}^i}^2+\vec{p}\,^2
+M^2(\omega_{f,n}^i,\vec{p}\,)\right]\right\}\\
&\quad-\dfrac{1}{2}\Bigg\{\sum_{f\in\{u,d,s\}}\left(\bar\sigma_f \bar S_f+\dfrac{G}{2}\bar S_f\bar S_f\right)+\dfrac{H}{2}\bar S_u\bar S_d\bar S_s\Bigg\}+\mathcal{U}(\Phi,\Phi^*,T)\,.
\end{aligned}
	\tag{\ref{OmegaMatsubara}$'$}
\end{equation*}
This is the thermodynamic potential of the nonlocal PNJL model in mean-field approximation. The auxiliary scalar fields $\bar S_f$ are determined by the SPA conditions, Eq.~\eqref{SPAbasis}.

\end{subsection}

\begin{subsection}{Gap equations in mean-field approximation}\label{MFgapssec}

Once the thermodynamic potential $\Omega$ is calculated, the fields $\sigma_u=\sigma_d,\sigma_s$ and $A_4^3,A_4^8$ can be determined by requiring thermodynamic potential to be stationary. The necessary conditions are given by the gap equations
\begin{equation}\label{thermalgaps}
\dfrac{\partial\Omega}{\partial\bar\sigma_u}=\dfrac{\partial\Omega}{\partial\bar\sigma_s}=\dfrac{\partial\Omega}{\partial A_4^3}=\dfrac{\partial\Omega}{\partial A_4^8}=0\ ,
\end{equation}
together with the stationary phase approximation equations \eqref{SPAbasis}.
  First, we limit ourselves to the zero-density, i.\,e.\ $\hat\mu=0$ case. Following Refs.~\cite{Ratti1,Simon1,Ratti2} we have then $\Phi=\Phi^*$ in the mean-field approximation and, consequently, $A_4^8=0$.

  Fig.~\ref{condPhicurrmass3f} shows the results for the temperature dependence of the chiral up- and strange-quark condensate and of the Polyakov-loop using the parameters given in Table~\ref{paramsI} (scenario~I) and Table~\ref{polyakovparameters}. This figure illustrates once more, as already demonstrated in Refs.~\cite{Hell,Simon1,Simon2}, the entanglement of chiral dynamics and Polyakov loop degrees of freedom, a characteristic feature of the PNJL approach. In the absence of a coupling between quark quasiparticles and Polyakov loop the chiral transition (for $N_\text{f}=3$ flavors) and the first-order deconfinement transition (of pure gauge QCD) appear at very different critical temperatures ($T_\text{chiral}\approx110\,\text{MeV}$ for  the chiral transition and $T_0\approx 270\,\text{MeV}$ for deconfinement). The presence of quarks breaks the $Z(3)$ symmetry explicitly and turns the first-order deconfinement phase transition into a continuous crossover. The quark coupling to the Polyakov loop moves this transition to lower temperature. At the same time the chiral transition (with explicit symmetry breaking by nonzero quark mass) turns into a crossover at an upward-shifted temperature, just so that both transitions nearly coincide at a common temperature $T_c\approx 200\,\text{MeV}$.

    \begin{figure}[!h]
\begin{center}
\includegraphics[width=.7\textwidth]{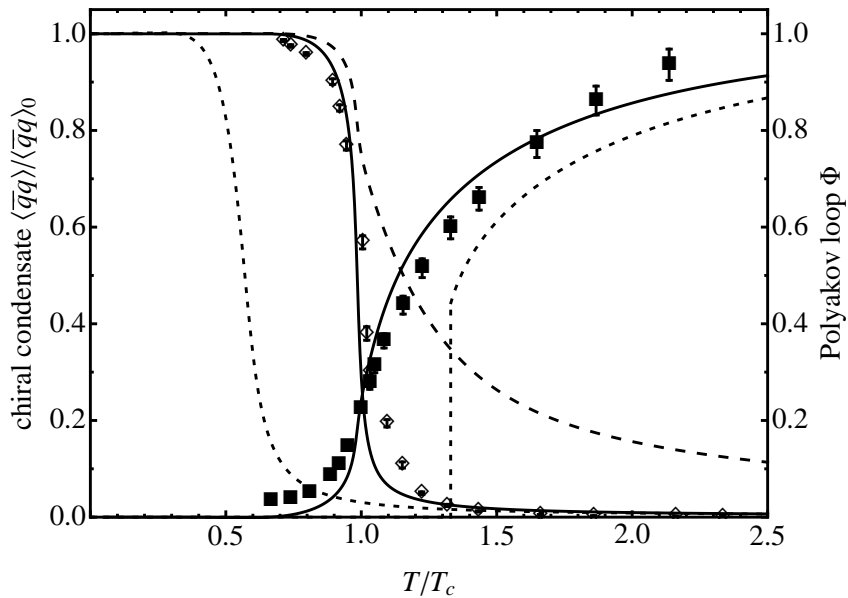}
\end{center}
\caption{Results of nonlocal PNJL calculations of the chiral and deconfinement transition pattern using the parameter set of scenario~I. Left solid curve: temperature dependence of the chiral condensate $\langle\bar uu\rangle=\langle\bar dd\rangle$. The strange-quark condensate $\langle\bar ss\rangle$ is shown as the dashed curve. Right solid curve: temperature dependence of the Polyakov loop $\Phi$. Left dotted curve: light quark condensate without coupling to Polyakov loop. Right dotted curve: Polyakov loop in the absence of quarks (pure gauge QCD). Also shown are lattice results for the chiral condensate and the Polyakov loop from the ``hotQCD'' collaboration \cite{Cheng}. The temperature is given in units of the transition temperature $T_c=200\,\text{MeV}$.}\label{condPhicurrmass3f}
\end{figure}

This symmetry breaking pattern seems also to be realized in recent lattice QCD results using staggered fermions \cite{Cheng} where a common chiral and deconfinement transition temperature $T_c=(196\pm3)\,\text{MeV}$ is observed. Fig.~\ref{condPhicurrmass3f} shows these lattice data for orientation. In the latest work of the same collaboration \cite{Cheng2}, domain-wall fermions are used instead of staggered fermions, leading to $T_c=171(10)(17)\,\text{MeV}$, a value which is consistent with both $T_c=196\,\text{MeV}$ and alternative lattice computations \cite{Aoki} that find a lower chiral transition temperature $T_c\simeq150\,\text{MeV}$ and a displacement from the deconfinement transition.

Compared to the $N_\text{f}=2$ flavor case (Ref.~\cite{Hell}) we observe only minor changes at this point, in particular the transition temperature decreases  slightly from $T_c^{2\,\text{f}}=207\,\text{MeV}$ to $T_c^{3\,\text{f}}=200\,\text{MeV}$. The transition temperature $T_c$ can be decreased further if the response from quark effects is included in the Polyakov-loop effective potential, leading to a lower $T_0$ in its parametrization. According to Ref.~\cite{Schaefer2} one has $T_0=190\,\text{MeV}$ for $2+1$ flavors instead of $T_0=270\,\text{MeV}$ for the pure gluon case.

\end{subsection}

\begin{subsection}{Parameter dependence}\label{comparison}

The chiral and deconfinement transition pattern shown in Fig.~\ref{condPhicurrmass3f} changes only marginally when scenario~I (with parameter set listed in Table~\ref{paramsI}) is replaced by scenario~II with a slightly larger coupling $G$. It is instructive also to examine the dependence on other parameters such as the current quark mass $m_u$. In the chiral limit, $m_u=m_d\to0$, the chiral condensate displays a second-order phase transition as expected. Explicit chiral symmetry breaking with $m_{u,d}\neq0$ turns this into a crossover transition as evident from Fig.~\ref{condPhicurrmass3f} for $m_{u,d}=3\,\text{MeV}$. Increasing the quark mass to $m_{u,d}=10\,\text{MeV}$ makes the crossover softer at $T>T_c$ while leaving the condensate unaltered at temperatures below $T_c$. The reason is that, below the transition temperature, the dynamical quark mass $M(p)$ entering the chiral condensate in the finite temperature generalization of Eq.~\eqref{chiralcondensate} is dominated by the large scalar field $\bar\sigma_u\simeq0.4\,\text{GeV}$. Changes of the light quark mass $m_{u,d}$ within a $10\,\text{MeV}$ range are not important compared to that scale, whereas they become more prominent above $T_c$ where $\bar\sigma_u$ drops rapidly. The softening of the strange-quark condensate $\langle\bar ss\rangle$ above $T_c$, as seen in Fig.~\ref{condPhicurrmass3f}, is much more pronounced, given the larger s-quark mass $m_s\simeq70\,\text{MeV}$.

Corrections to the behavior of the chiral condensate and the pressure below $T_c$ come primarily from thermal pions (and kaons) as will be discussed in the following subsection.

\end{subsection}

\begin{subsection}{Beyond mean field: mesonic corrections}\label{rpa}

So far the calculations have been performed in the mean-field approximation in  which the pressure $P=-\Omega$ is determined by the quarks moving as quasiparticles in the background provided by the expectation values of the sigma fields, $\bar\sigma_u=\bar\sigma_d$ and $\bar\sigma_s$, and of the Polyakov loop $\Phi$. In order to get a realistic description of the hadronic phase (at temperatures $T\lesssim T_c$), it is important to include mesonic quark-antiquark excitations. The hadronic phase in the absence of baryons is dominated by (the light) pseudoscalar mesons (pions and kaons). The quark-antiquark continuum is suppressed by confinement. However, as pointed out  in Ref.~\cite{Hansen}, mesons described within the standard (local) PNJL model can still undergo unphysical decays into the quark-antiquark continuum even below $T_c$. In the nonlocal PNJL model, such unphysical decays do not appear by virtue of the momentum-dependent dynamical quark mass\footnote{With the possible exception of the $\eta'$-meson which will not be considered in this section.}.  This means that the pressure below $T_c$ is basically generated by the pion and kaon poles of the corresponding one-loop $q\bar q$ Green functions, with their almost temperature independent position. Therefore the calculated pressure below $T_c$ corresponds to that of a boson gas with constant masses.

To include the mesonic contributions to the pressure in our nonlocal PNJL model we can basically use the formalism described in Ref.~\cite{Hansen}. One has to calculate $G_{\text{PS},\text{S}}(\nu_m,\vec{p}\,)$, Eq.~\eqref{G}, and, in particular, the quark loop contribution to the pseudoscalar (PS) and scalar (S) mesonic self-energies $\varPi_{\text{PS},\text{S}}(\nu_m,\vec{p}\,)$ (where $\nu_m=2\pi m T, m\in\Z$ is the bosonic Matsubara frequency and $\vec{p}$ is the momentum of the incoming meson), given in Eq.~\eqref{varPi}, at finite temperature. Using the replacement $\int\frac{\diff^4 p}{(2\pi)^4}\to T\sum_{n\in\Z}\int\frac{\diff^3 p}{(2\pi)^3}$  and the rules \eqref{gauging} this can be carried out easily. The additional contribution of mesonic quark-antiquark modes to the pressure is given by a ring sum of random phase approximation (RPA) chains, investigated in Ref.~\cite{Huefner} and leading to the expression
\begin{equation}\label{mesonpress}
 	P_\text{meson}(T)=-T\sum_{M=\text{PS},\text{S}}\dfrac{d_M}{2} \sum_{m\in \Z}\int\dfrac{\diff^3 p}{(2\pi)^3}\ln\left[G_M(\nu_m,\vec{p}\,)\right],
\end{equation}
where $d_M$ is the mesonic degeneracy factor ($d_M=3$ for pionic and $d_M=4$ for kaonic modes).
Because of the momentum dependence of the nonlocality distribution $\mathcal{C}(p)$ and the dynamical quark masses $M_q(p)$, integrations and summations in Eq.~\eqref{mesonpress} can only be carried out numerically. 

Results for the pressure in the presence of pion, kaon and scalar modes are presented in Fig.~\ref{pressplot}.\footnote{The plot shown in Fig.~\ref{pressplot} uses the parameters of scenario~I. The difference between pressure curves calculated from parameter sets I and II turns out to be negligibly small.} Apart from the full result (solid line) we show the mean-field result (MF, with the pressure determined by quark quasiparticles only) and the mean-field result plus pion and corresponding scalar contributions. It is evident that at low temperatures the mean-field contribution from the quarks is suppressed and the pressure can be described by a free meson gas. Near the transition temperature the scalar mesonic modes give a small additional contribution. Finally, above temperatures $T>1.5\,T_c$ the mesonic contributions become negligible and the quark-gluon mean fields dominate the pressure.

\begin{figure}
\begin{center}
\includegraphics[width=.7\textwidth]{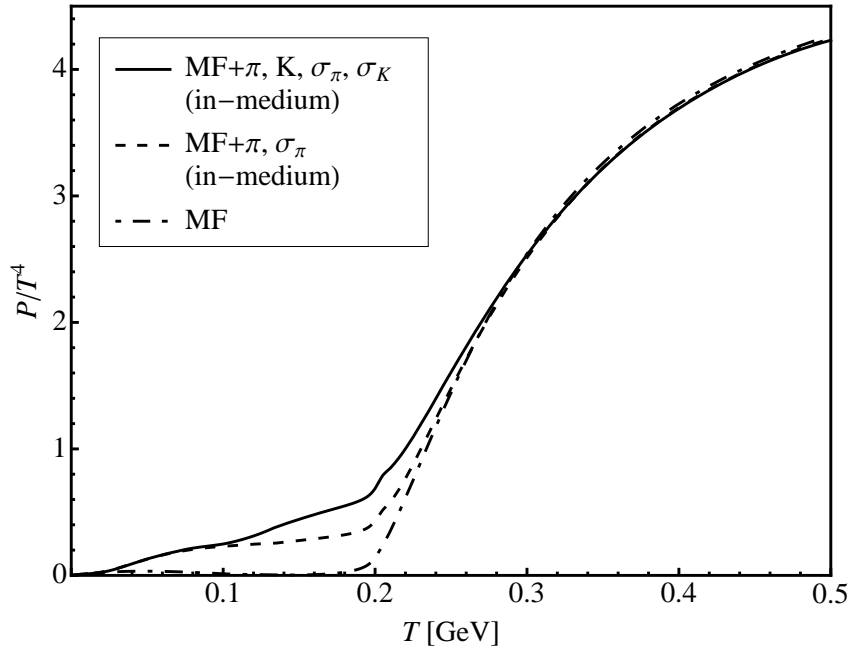}
\caption{Pressure (in units of $T^4$) calculated in the nonlocal PNJL model as a function of  temperature on an absolute temperature scale. Solid curve: full calculation (i.\,e.~mean-field result plus mesonic corrections). Dash-dotted curve: mean-field result (no mesonic corrections). Dashed curve: mean-field plus pionic and corresponding scalar modes.\label{pressplot}}
\end{center}
\end{figure}

The RPA treatment of mesonic contributions to the pressure allows us, in addition, to get the corrections to the chiral condensates. Exploiting the definition of the chiral condensate,
	\begin{equation}
	 	\langle \bar q q\rangle=\dfrac{\int\mathscr{D} A\mathscr{D}\bar\psi\mathscr{D}\psi\, \bar qq\,\euler^{-\mathcal{S}_\text{E}}}{\int\mathscr{D} A\mathscr{D}\bar\psi\mathscr{D}\psi\,\euler^{-\mathcal{S}_\text{E}}}=\dfrac{\partial\Omega}{\partial m_q}\ .
	\end{equation}
The pionic corrections to $\langle\bar uu\rangle$ is computed by differentiating the pion pressure of Eq.~\eqref{mesonpress} with respect to the up-quark current quark mass:
	\begin{equation}
	 	\delta_\pi\langle\bar uu\rangle=-\dfrac{\partial P_\pi}{\partial m_u}.
	\end{equation}
It turns out, as expected, that the modification of the chiral condensate owing to pions is very similar to the results from chiral perturbation theory (cf., e.\,g.,~Ref.~\cite{chipt}).  At temperatures below $T_c$ pions tend to soften the condensate and make the chiral transition smoother in the range $0.5 T_c<T<T_c$.

Further quantities of interest are the energy density $\varepsilon$ and the trace anomaly, $(\varepsilon- 3P)/T^4$. The trace anomaly, in particular, is relevant here since it is the quantity which can be directly computed in lattice simulations (Ref.~\cite{Cheng}). This quantity, representing the trace of the energy-momentum tensor, is the ``interaction'' measure which can be expressed in terms of a derivative of the pressure with respect to temperature:
\begin{equation*}
 	\dfrac{\varepsilon-3 P}{T^4}=T\dfrac{\partial}{\partial T}\left(\dfrac{P}{T^4}\right).
\end{equation*}
From there it is straightforward to calculate the energy density.
Figs.~\ref{conformalmeasure} and \ref{energydensity} show the results that follow from Fig.~\ref{pressplot} both for the mean-field case and with the additional inclusion of mesonic (pionic and kaonic) contributions. Our results are compared to 3-flavor lattice data of Ref.~\cite{Cheng}.

\begin{figure}[t]
\begin{center}
	\begin{minipage}[t]{.475\textwidth}{
		\includegraphics[width=\textwidth]{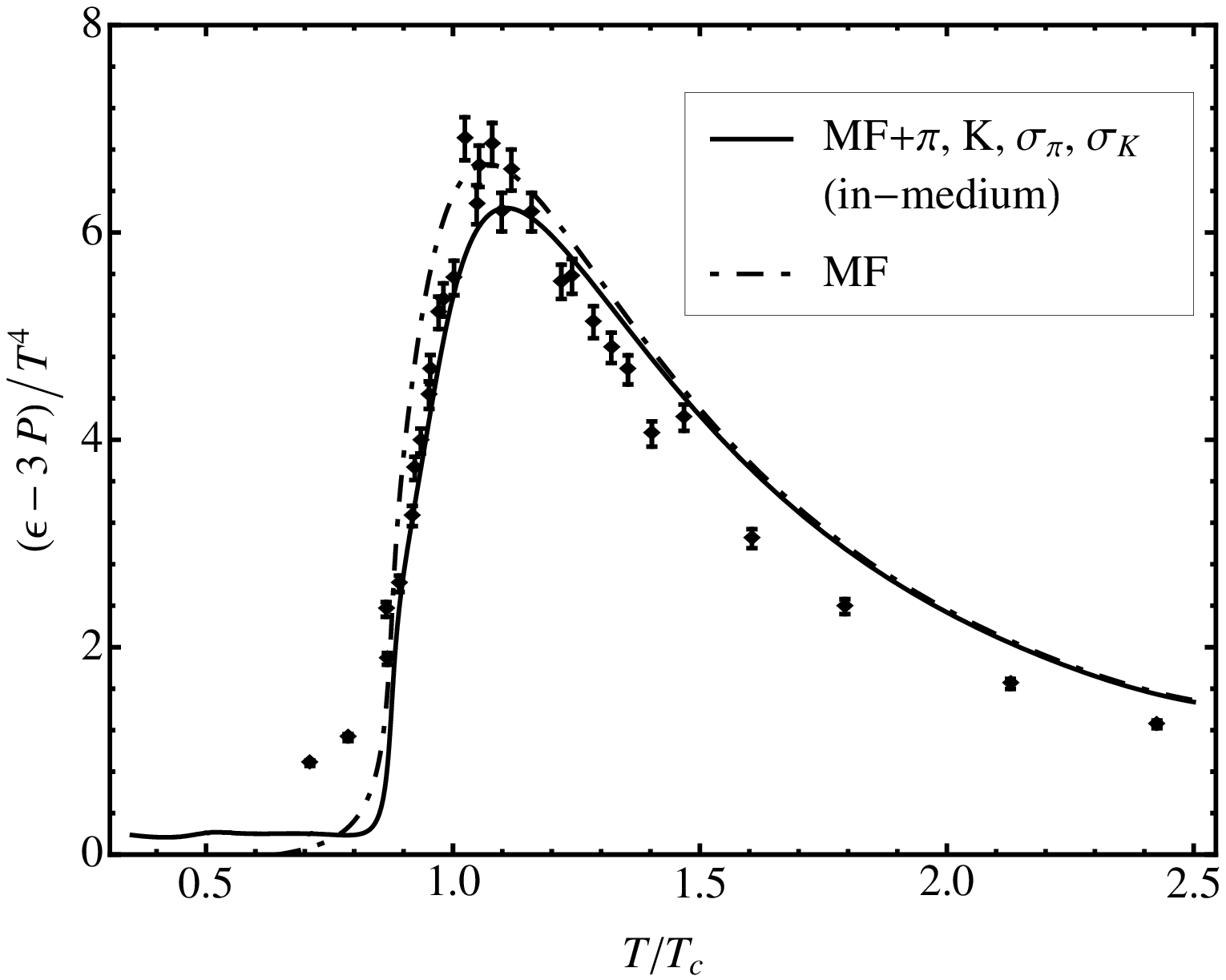}
		\caption{Trace anomaly $(\varepsilon-3P)/T^4$. Dash-dotted curve: mean-field results. Solid curve: full calculation with inclusion of pions, kaons and scalar modes (results for scenario~I). For comparison, 3-flavor lattice results from Ref.~\cite{Cheng} are shown.\label{conformalmeasure}}}
			
	\end{minipage}
\hfill
	\begin{minipage}[t]{.475\textwidth}{
		\includegraphics[width=\textwidth]{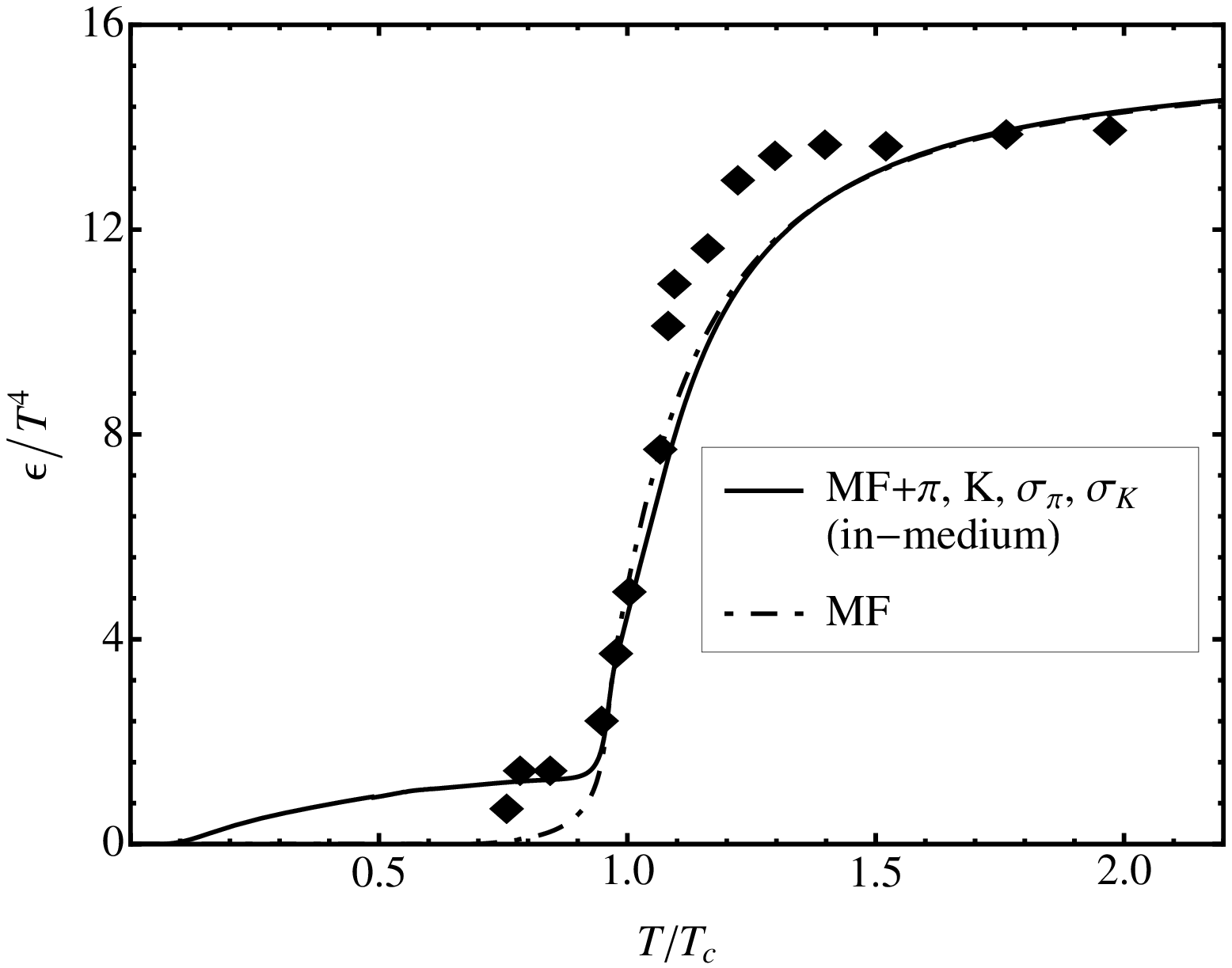}
		\caption{Energy density $\varepsilon/T^4$ for the mean-field and RPA-case. (Legend as in Fig.~\ref{conformalmeasure}.) Lattice data are borrowed from Ref.~\cite{Cheng}.\label{energydensity}}}
	\end{minipage}
\end{center}
\end{figure}

\end{subsection}

\begin{subsection}{PNJL thermodynamics at finite quark chemical potential}\label{finitedensity}

The nonlocal PNJL approach described in this work can be extended to finite quark chemical potential $\hat\mu=\text{diag}(\mu_u,\mu_d,\mu_s)$. We do this here with the aim of drawing a schematic phase diagram in the $(T,\mu_u)$ plane. We set $\hat\mu=\text{diag}(\mu,\mu,0)$ (i.\,e. we work in the isospin symmetric case $\mu_u=\mu_d$ and with $\mu_s=0$).  For the sake of simplicity we restrict ourselves to a scenario without diquark condensates.

The introduction of a chemical potential\footnote{For convenience we use a quark chemical potential throughout this work. The corresponding baryon chemical potential is three times the u-quark chemical potential, i.\,e.~$\mu_\text{B}=3\mu_u$.} is accomplished using the  prescriptions of the Matsubara formalism (see Ref.~\cite{Kapusta} and Eq.~\eqref{nambugorkovS}):  shift the frequencies $\omega_{f,n}\to\omega_{f,n}-\imu\mu_f$ in the particle sector (i.\,e. in the upper-left submatrix) of the Nambu-Gor'kov propagator \eqref{nambugorkovS} and  replace $\omega_{f,n}\to\omega_{f,n}+\imu\mu_f$ in the corresponding antiparticle sector (i.\,e. the lower-right submatrix). It is then straightforward to compute the thermodynamic potential $\Omega(T,\hat\mu)$ at nonzero $\hat\mu$, following Eqs.~\eqref{OmegaMatsubara}, \eqref{nambugorkovS} with Matsubara frequencies properly shifted by the chemical potential.

We focus here on the $T$ and $\mu_u$ dependence of the scalar field $\bar\sigma_u$ that acts as a chiral order parameter, deduced from the condition $\frac{\partial\Omega(T,\mu)}{\partial\bar\sigma}=0$. The results are shown in Figs.~\ref{phasediag3f_sig360} and \ref{phasediag3f} for scenarios~I and II, respectively. The profile of $\bar\sigma_u$ displays once again the chiral crossover transition at $\mu_u=0$. It turns into a first-order phase transition at a critical point (located at $T_\text{CEP}\approx170\,\text{MeV}$ and $\mu_\text{CEP}\approx180\,\text{MeV}$  for scenario~I and at $T_\text{CEP}\approx195\,\text{MeV}$ and $\mu_\text{CEP}\approx110\,\text{MeV}$ for scenario~II). This qualitative feature is typical for NJL or PNJL type models with or without explicit diquark degrees of freedom (see e.\,g.~Refs.~\cite{Ratti1,Simon1,Buballa}). Related work is reported in Refs.~\cite{Sasaki,Abuki} where the nonlocality of the fermionic interaction is introduced only in the three-momentum sector. The critical points found in the present calculations differ both in their $T$ and $\mu_u$ values from the aforementioned references. Reasons for such differences are given in Refs.~\cite{Simon1,Fukushima3} where it is pointed out and demonstrated that the location of the critical point is extremely sensitive to model details and input parameters.

    \begin{figure}[t]
\begin{center}
\begin{minipage}[t]{.475\textwidth}{
\includegraphics[width=\textwidth]{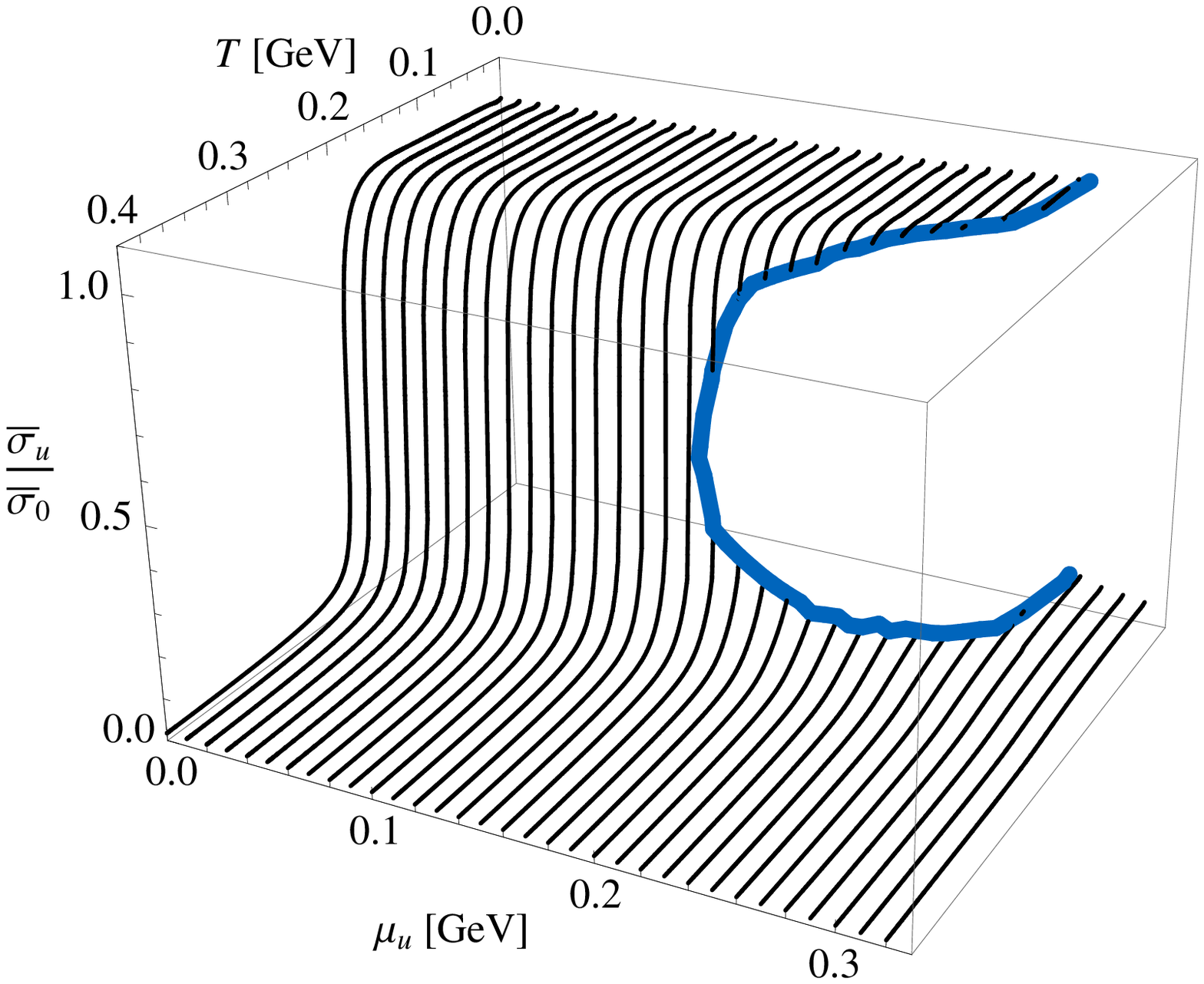}
\caption{Chiral order parameter $\bar\sigma_u$ for scenario~I, normalized to its value $\bar\sigma_0$ at $T=\mu=0$, as a function of temperature and u-quark chemical  potential (note $\mu_u=\mu_d,\mu_s=0$). The thick blue line shows the border for the first-order transition.}\label{phasediag3f_sig360}
}
\end{minipage}
\hfill
\begin{minipage}[t]{.475\textwidth}{
\includegraphics[width=\textwidth]{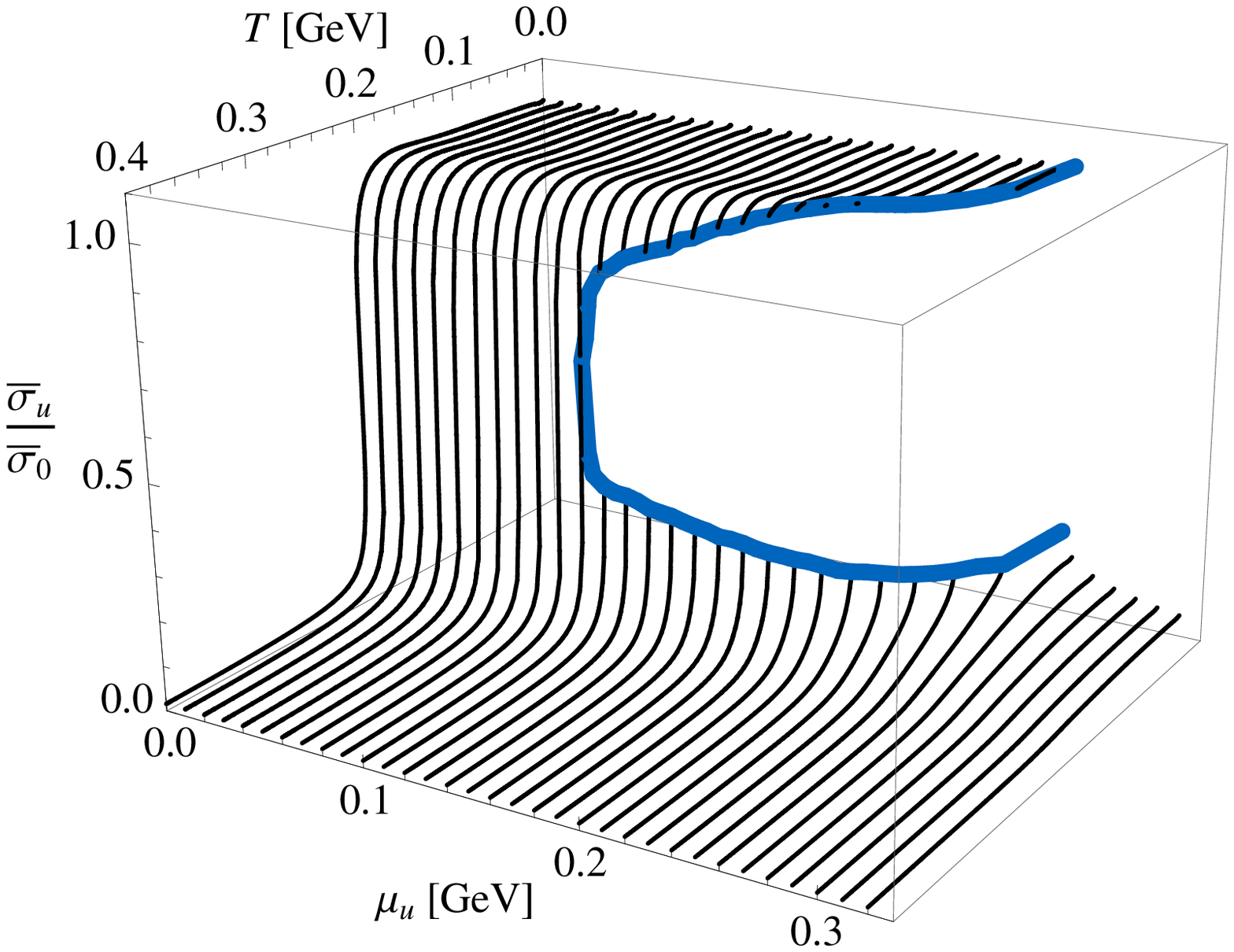}
\caption{Same as in Fig.~\ref{phasediag3f_sig360}, for parameters of scenario~II. The major difference between Figs.~\ref{phasediag3f_sig360} and \ref{phasediag3f} is the location of the critical point:   $(\mu_\text{CEP},T_\text{CEP})^\text{(I)}=(180\,\text{MeV},170\,\text{MeV})$ compared to $(\mu_\text{CEP},T_\text{CEP})^\text{(II)}=(115\,\text{MeV},195\,\text{MeV})$.}\label{phasediag3f}
}
\end{minipage}
\end{center}
\end{figure}

The projection of Figs.~\ref{phasediag3f_sig360} and \ref{phasediag3f} onto the $T$-$\mu_u$ plane gives the phase diagram of the nonlocal PNJL model, Figs.~\ref{quarkyonic}. 
At low $\mu$ this phase diagram shows the chiral and deconfinement crossover transitions in close contact as already discussed. The deconfinement transition is displayed here as a band bounded by dashed lines where the upper and lower bounds are given by values $\Phi=0.5$ and $\Phi=0.3$ of the Polyakov loop, respectively, reflecting the relatively soft crossover of this transition (see also Fig.~\ref{condPhicurrmass3f}). At larger values of the chemical potential, beyond the critical point, a separation between the chiral and deconfinement transition takes place.

    \begin{figure}[t]
\begin{center}
	\begin{minipage}[t]{.475\textwidth}{
\includegraphics[width=\textwidth]{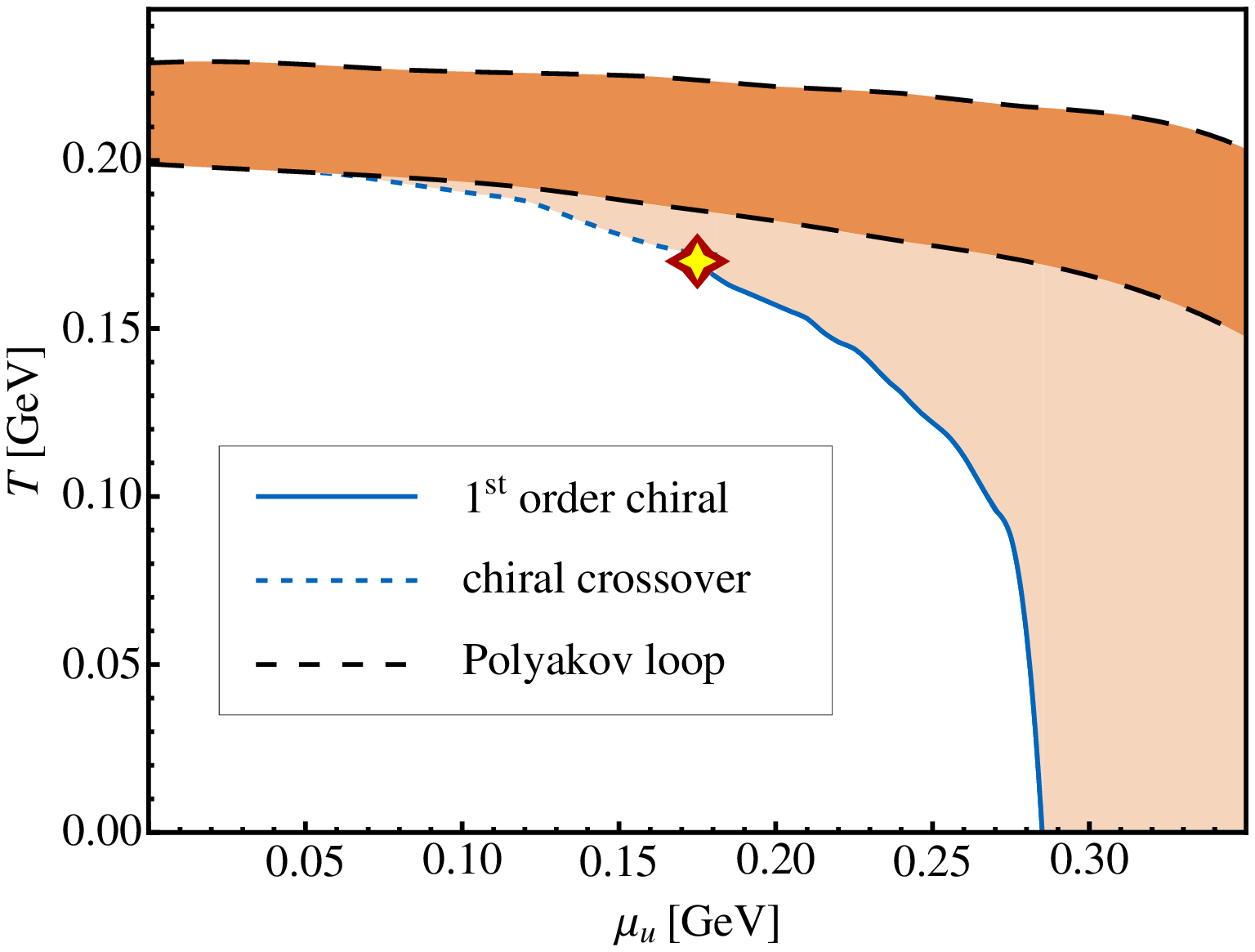}
	}
	\end{minipage}
\hfill
	\begin{minipage}[t]{.475\textwidth}{
\includegraphics[width=\textwidth]{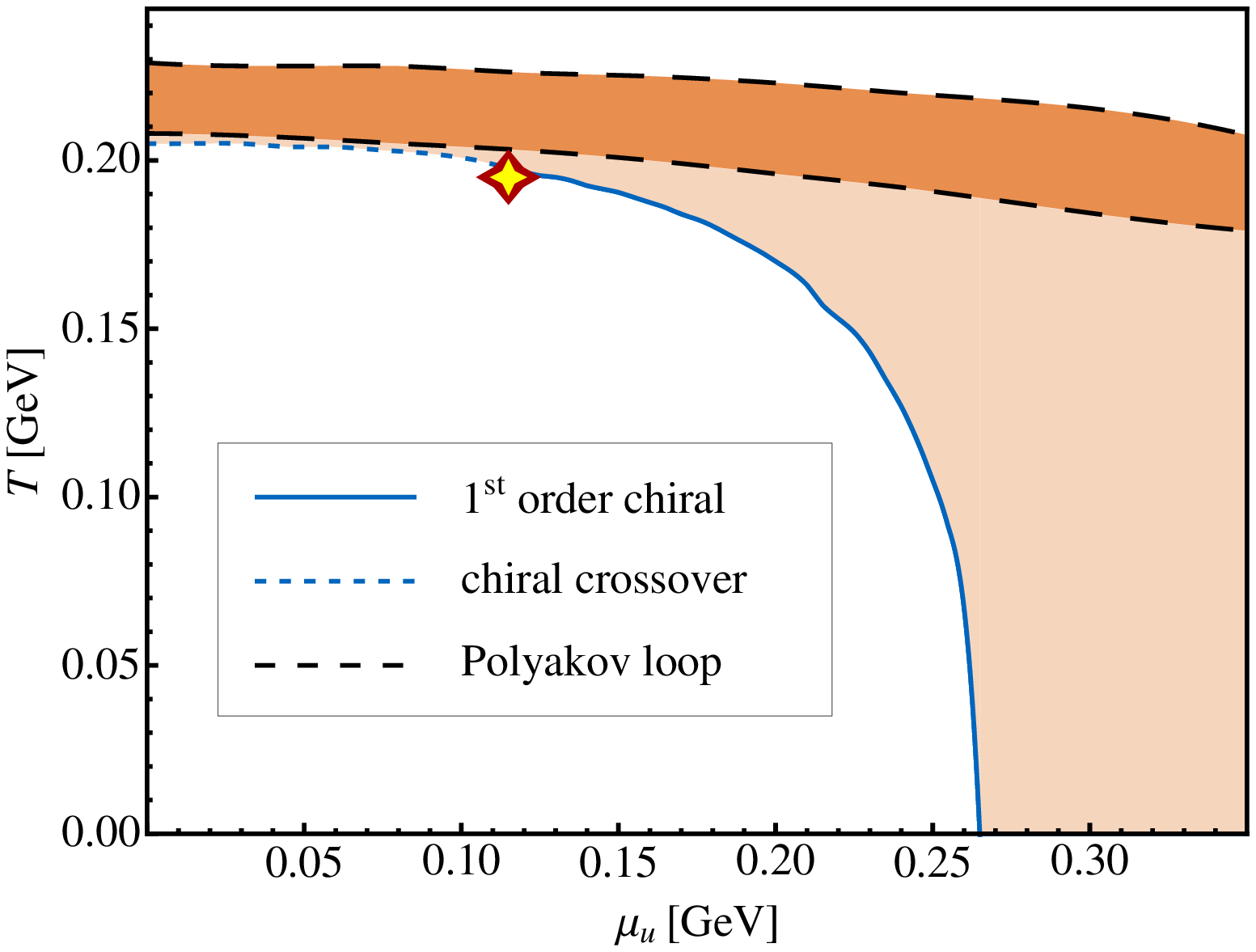}
	}
	\end{minipage}
\caption{Phase diagram calculated within the nonlocal PNJL model using the parameters of scenario~I (left picture) and scenario~II (right picture). The solid blue line shows the first-order chiral transition (the star denotes the critical end point). The short-dashed blue line marks the (chiral) crossover transition while the long-dashed black lines correspond to the deconfinement transition (the lower and upper lines correspond to  $\Phi=0.3$ and $\Phi=0.5$, respectively).}\label{quarkyonic}
\end{center}
\end{figure}

The area between the (first-order) chiral phase transition and the deconfinement crossover has recently been interpreted in terms of a ``quarkyonic'' phase \cite{quarkyoniclit}. It should be pointed out, however, that, at nonzero baryon densities and low temperature, PNJL type models are only schematic and cannot be considered as realistic.  From Fig.~\ref{quarkyonic} it appears that the chiral first-order transition boundary meets the $\mu$ axis at $T=0$ for values of the baryon chemical potential as small as $\mu_B=3\mu<0.9\,\text{GeV}$. This is the domain of nuclear matter that is known to be a Fermi liquid of nucleons. The PNJL model works instead with quarks as quasiparticles, the ``wrong'' degrees of freedom in this low-temperature phase at moderate baryon densities.

The deconfinement transition band at large $\mu$ has been calculated using the Polyakov-loop effective potential \eqref{polyakovU}. This effective potential does not include higher order effects due to the presence of quarks at nonzero chemical potential. Such additional $\mu$-dependent effects are expected to move the crossover boundary to lower temperatures as $\mu$ increases \cite{Pawlowski,Braun}.

Fig.~\ref{quarkyonic} demonstrates that the position of the critical point is sensitive to small changes of the four-fermion coupling $G$. The relatively small increase of this coupling strength between scenarios~I and II, keeping hadronic vacuum properties at $T=\mu=0$ almost unchanged, results nevertheless in a significant shift of the critical point in the $(T,\mu)$ plane.

We also confirm that the phase structure is sensitive to variations of the 't~Hooft interaction coupling strength $H$ (see Fig.~\ref{critpoint}). An increase of  $H$ by only $5\,\%$ turns the chiral transition even at $\mu_u=0$ into a first-order transition. This behavior might be expected considering the Columbia plot (Ref.~\cite{Brown}). On the other hand, a decrease of $H$, corresponding to a reduced $\eta'$ mass in the thermal medium, moves the end point to higher chemical potentials and lower temperatures. The sensitivity to the axial anomaly observed here in the nonlocal PNJL model is, however, less pronounced than that in the local model Ref.~\cite{Fukushima3,Bratovic}. We do not observe that the end point is removed altogether from the phase diagram as quickly as in the local PNJL model.

Finally, using Eq.~\eqref{omegasimpel} we calculate the pressure $P=-\Omega$ at finite chemical potential. In Fig.~\ref{DeltaPress_selection_sig360} the (normalized) pressure difference
	\begin{equation}\Delta P(T,\mu_u):=P(T,\mu_u)-P(T,\mu_u=0)\end{equation}
is shown for selected values of $\mu_u=\mu_d$ and compared to (2-flavor) lattice data from Ref.~\cite{Allton}. Fig.~\ref{DeltaPress_sig360} displays the full result in the $T$-$\mu_u$ plane. Both figures have been obtained using the parameter set of scenario~I.

As already mentioned, the pictures drawn in Figs.~\ref{phasediag3f_sig360}, \ref{phasediag3f} and \ref{quarkyonic} are to be taken as only schematic, for several reasons. First, the location of the critical point is  sensitive not only to the coupling strengths $G$ and $H$, but also to the input current quark mass \cite{Simon1}. Second, the almost constant behavior of $\bar\sigma_u(T=0,\mu_u)$ with increasing quark chemical potential is unrealistic in the absence of explicit baryon (nucleon) degrees of freedom including their interactions.

\begin{figure}[!t]
\begin{center}
	\begin{minipage}[t]{.475\textwidth}{
\includegraphics[width=\textwidth]{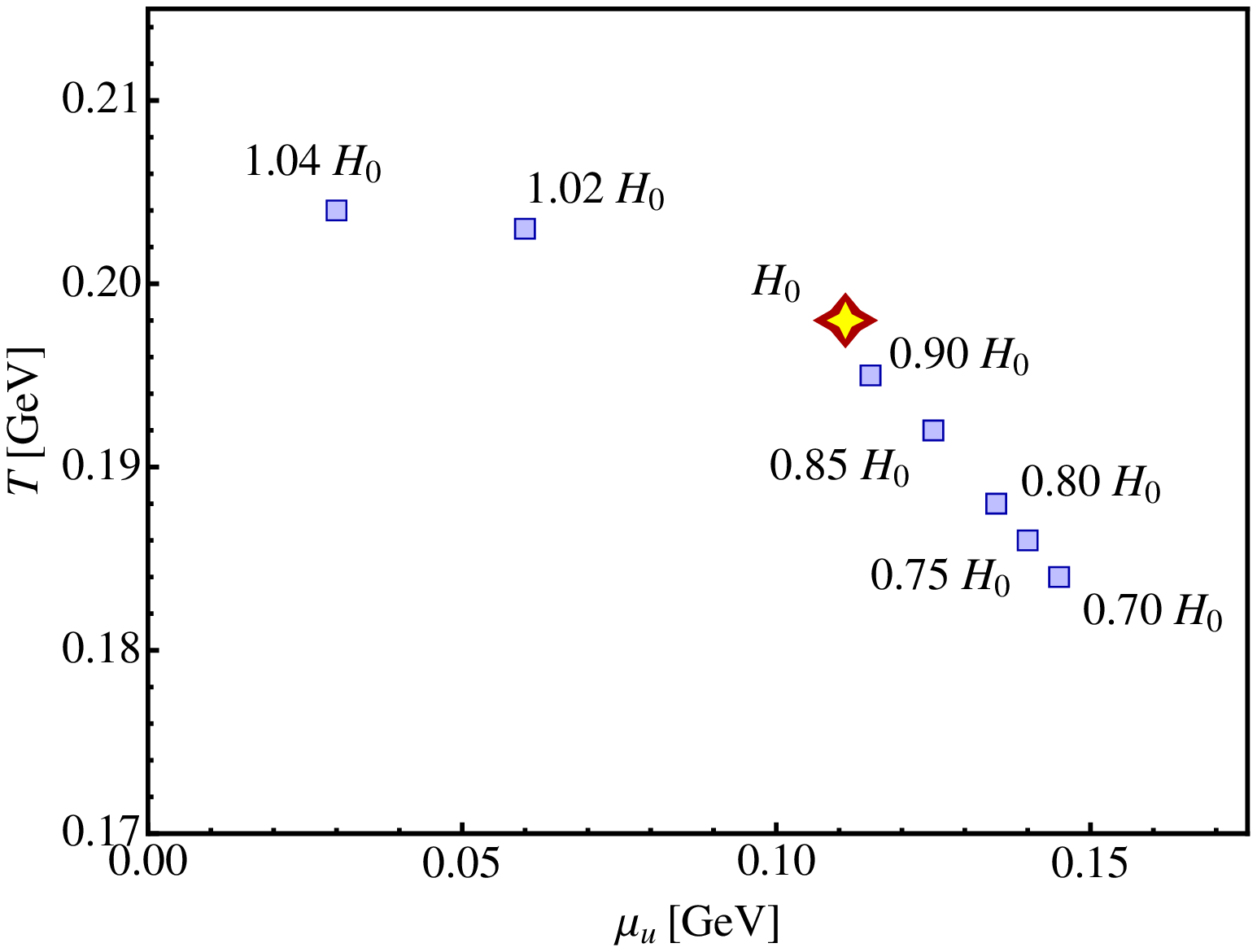}
\caption{Location of the critical point depending on several values of the 't~Hooft coupling strength $H$ in units of the coupling strength $H_0$ of scenario~II.}\label{critpoint}	}
	\end{minipage}
\hfill
	\begin{minipage}[t]{.475\textwidth}{
\includegraphics[width=\textwidth]{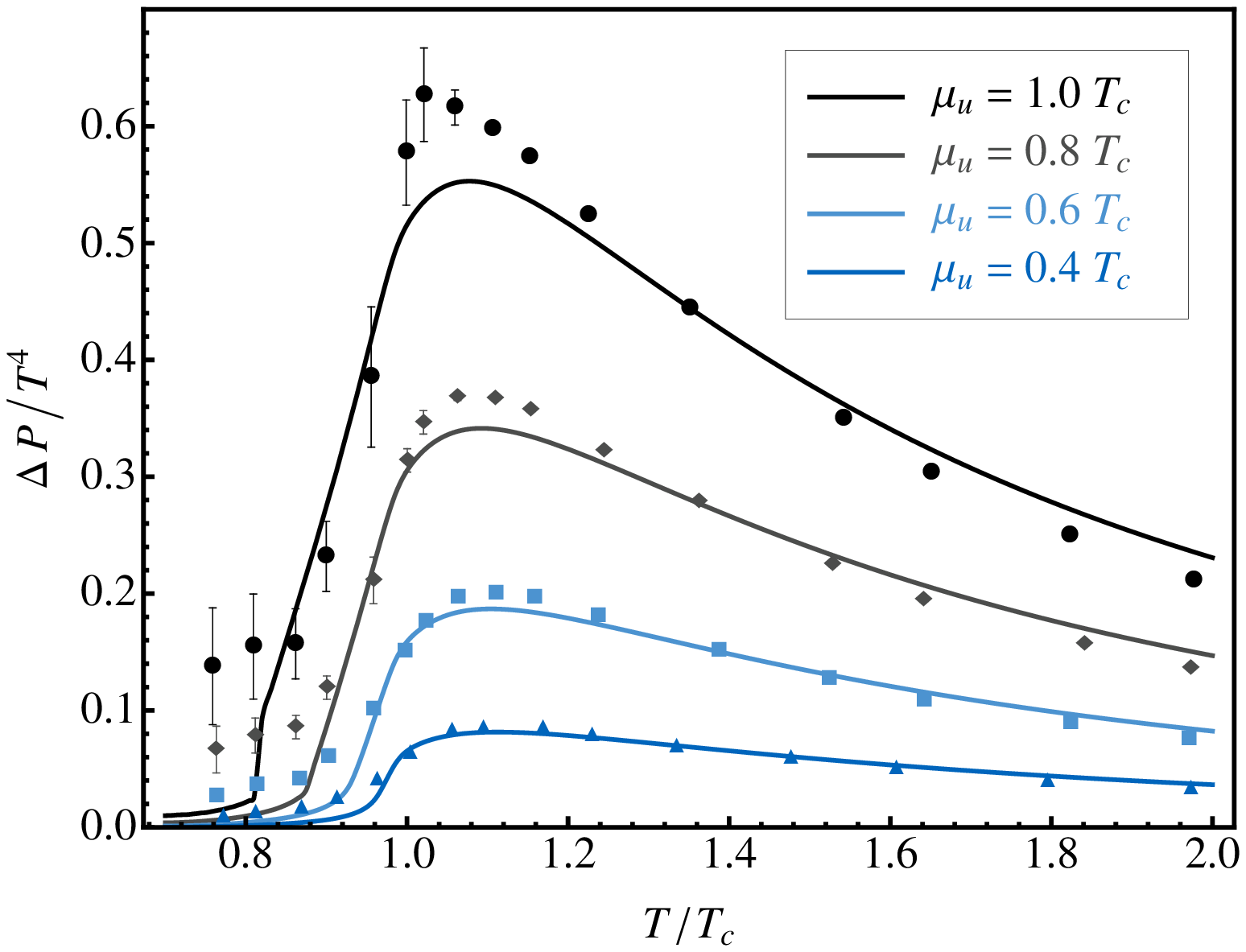}
\caption{Pressure difference at finite chemical u-quark potential, $\Delta P(T,\mu_u)=P(T,\mu_u)-P(T,0)$ (parameters of scenario~I) for selected values of $\mu_u$ compared to lattice data from Ref.~\cite{Allton}.}\label{DeltaPress_selection_sig360}
	}
	\end{minipage}
\end{center}
\begin{center}
\includegraphics[width=.7\textwidth]{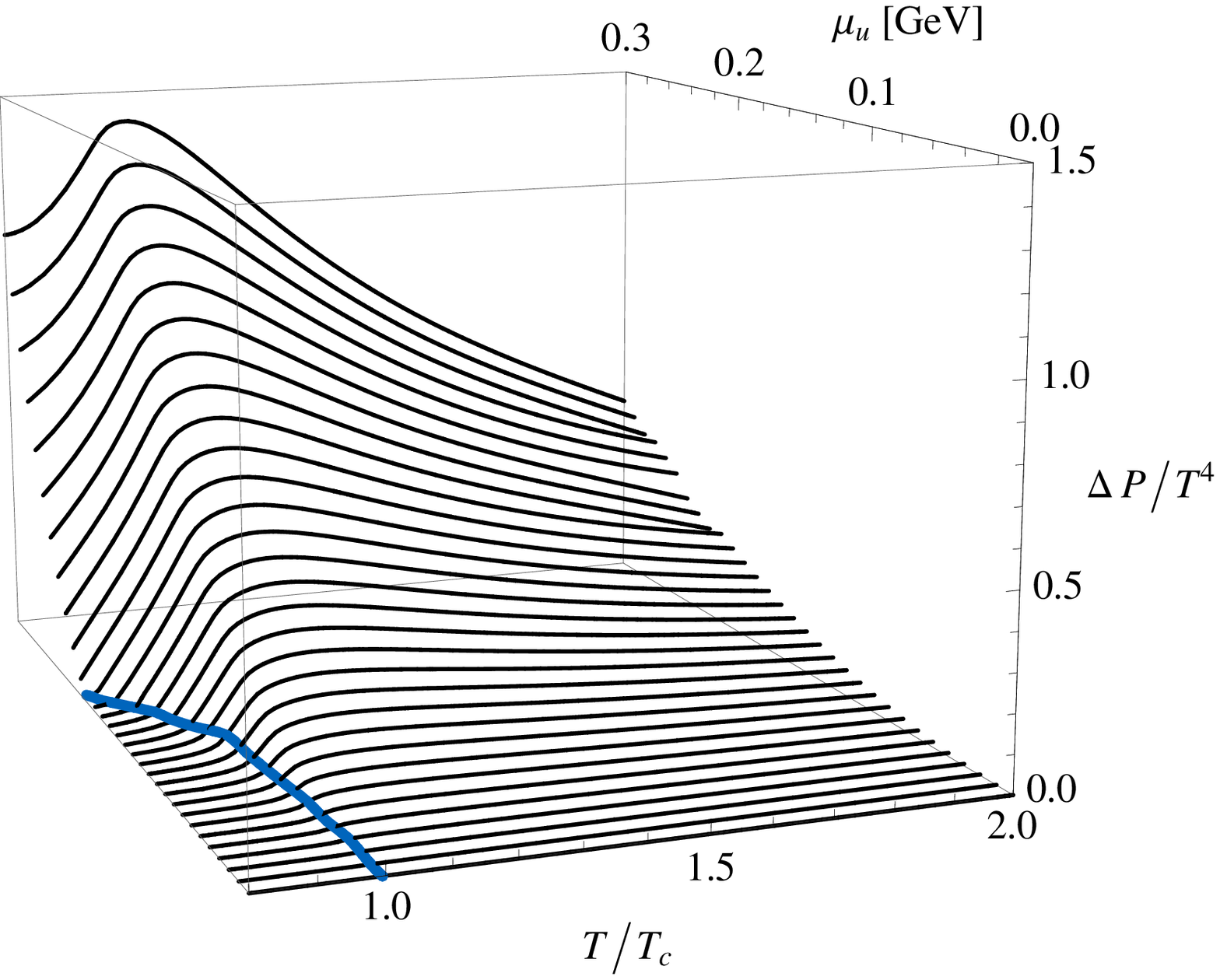}
\caption{Excess pressure $\Delta P(T,\mu_u)=P(T,\mu_u)-P(T,0)$ at finite chemical u-quark potential calculated in the nonlocal PNJL model in units of $T^4$ for scenario~I. The thick solid blue line indicates the chiral crossover and first-order transition line. \label{DeltaPress_sig360}}
\end{center}
\end{figure}

What is actually required as a starting point for extensions to nonzero chemical potential is a realistic equation of state at finite baryon density, incorporating the known properties of equilibrium and compressed nuclear matter. In such a framework \cite{Kaiser}, the density dependence of the chiral condensate $\langle\bar u u\rangle$ (or of the scalar field $\bar\sigma_u$) is well-known to be quite different from the profile shown in Figs.~\ref{phasediag3f_sig360}, \ref{phasediag3f}. The magnitude of $\langle\bar uu\rangle$ decreases linearly with density $\rho$ \cite{densdep}, with a slope controlled by the pion-nucleon sigma term, and then stabilizes at densities above normal nuclear matter through a combination of two- and three-body correlations and Pauli blocking effects. The transition towards chiral symmetry restoration at $T=0$ is shifted beyond at least twice the density of normal nuclear matter \cite{Kaiser}.

Irrespective of these comments, the nonlocal PNJL approach is obviously instructive in modeling the chiral and deconfinement thermodynamics at $\hat\mu=0$. Dealing with finite baryon density requires ultimately yet another synthesis, namely a matching of PNJL above the chiral transition and in-medium chiral effective field theory with baryons below that transition.

\end{subsection}

\end{section}

\begin{section}{Conclusions and outlook}\label{summary}

We summarize our findings as follows:

i) The nonlocal generalization of the PNJL model to $N_\text{f}=3$ flavors incorporates all important nonperturbative features of low-energy QCD with inclusion of strange quarks: spontaneous and explicit chiral symmetry breaking, the axial anomaly, and thermodynamical aspects of confinement in terms of the Polyakov loop. A separable form of the underlying nonlocal effective interactions between quarks, including the axial $\text{U}(1)_\text{A}$ breaking six-fermion vertex, proves to be successful in reproducing the physics of the pseudoscalar meson nonet. Chiral low-energy theorems are shown to be fulfilled.

ii) At the level of generalized gap equations, the resulting momentum-dependent quark (quasiparticle) masses permit establishing connections with instanton physics, lattice QCD and (Landau gauge) Schwinger-Dyson approaches. The nonlocality distribution of the effective interaction between quarks reflects typical instanton sizes of about $1/3\,\text{fm}$. This distribution replaces the artificial sharp momentum space cutoff in standard (local) NJL type models and links the quark mass function $M(p)$ smoothly to the correct QCD behavior at large momentum scales.

iii) The thermodynamics of the three-flavor nonlocal PNJL model reproduces corresponding $N_\text{f}=2+1$ lattice QCD results with almost physical quark masses surprisingly well. In particular, the dynamical entanglement of the chiral and deconfinement crossover transitions observed previously in two-flavor PNJL models is confirmed also for $N_\text{f}=3$. The interaction measure $\varepsilon-3 P$ is well reproduced around the critical temperature $T_c\simeq 0.2\,\text{GeV}$. Mesonic contributions to the pressure can be systematically incorporated.

iv) The critical  point in the phase diagram and its location in the plane of temperature $T$ and baryon chemical potential $\mu_\text{B}$ remain an open issue. The position of this critical point turns out to be extremely sensitive to fine-tunings of parameters (quark masses, coupling constants), as already found in previous studies.

v) Previous calculations at large quark chemical potentials, using local PNJL models, were limited by the momentum space cutoffs characteristic of such models. While these limitations are overcome in the nonlocal PNJL approach, one must still be aware of the fact that, at low temperatures and moderate baryon densities, PNJL models do not operate with the proper nucleon degrees of freedom relevant at such densities. In the baryonic phase around and below $\mu_\text{B}\sim1\,\text{GeV}$, the phase diagram should represent a nuclear Fermi liquid and not a quarkyonic quasiparticle system.

Implementing the constraints from realistic nuclear or neutron matter equations of state on the QCD phase diagram remains a challenge for the future.

\end{section}

\begin{section}*{Acknowledgements}
 
Stimulating discussions and communications with Kenji Fukushima and Norberto Scoccola are gratefully acknowledged.

\end{section}

\newpage
\begin{appendix}
\begin{section}{Derivation of the 't~Hooft interaction}\label{hooftappendix}

In this appendix we show how the Kobayashi-Maskawa-'t~Hooft determinant expression, Eq.~\eqref{thooft}, can be cast into the form used in this work, Eq.~\eqref{sint6sep}.

In order to write the 't\,Hooft determinant in a more tractable way, we apply Newton's and Girard's formula
\begin{equation}\label{newton}	\det\mathcal{J}^\pm=\dfrac{1}{6}\left(\text{tr}\mathcal{J}^\pm\right)^3-\dfrac{1}{2}\left(\text{tr}\mathcal{J}^\pm\right)\left(\text{tr}{\mathcal{J}^\pm}^2\right)+\dfrac{1}{3}\text{tr}{\mathcal{J}^\pm}^3.
\end{equation}
Here $\text{tr}$ indicates the trace over flavor space only.
We use the Gell-Mann matrices as a basis in flavor space, $\{\lambda_0,\lambda_1,\dots,\lambda_8\}$, with the additional definition $\lambda_0:=\sqrt{\frac{2}{3}}\,\text{diag}(1,1,1)$ in order to maintain 	$\text{tr}\{\lambda_\alpha\cdot\lambda_\beta\}=2\delta_{\alpha\beta}$ for all $\alpha,\beta\in\{0,\dots,8\}$. This allows us to write
\begin{equation}\label{chooft}
 	\mathcal{J}^\pm=\sum_{\alpha=0}^8 c_\alpha^\pm \lambda_\alpha\qquad\Longleftrightarrow\qquad \text{tr}\{\lambda_\alpha\mathcal{J}^\pm\}=2 c_\alpha^\pm
\end{equation}
and, consequently, $c_\alpha^\pm=\frac{1}{2}\text{tr}\{\lambda_\alpha\mathcal{J}^\pm\}$.

Furthermore, from Eq.\,\eqref{thooft} we have with the definitions \eqref{j}
\begin{align*}
 	\dfrac{1}{2}\text{tr}\{\lambda_\alpha\mathcal{J}^\pm(x)\}&=\dfrac{1}{4}\int\diff^4 z\,\lambda_\alpha^{ij}\bar\psi_i\!\left(x+\frac{z}{2}\right)(1\mp\gamma_5)\,\mathcal{C}(z)\,\psi_j\!\left(x-\frac{z}{2}\right)\\
		&=\dfrac{1}{4}j_\alpha^S(x)\mp\dfrac{1}{4\,\imu} j_\alpha^P(x).
\end{align*}
By means of Eq.\,\eqref{chooft}, this allows to write $c_\alpha^\pm=\frac{1}{4}j_\alpha^S\pm\frac{\imu}{4}j_\alpha^P$, or, inversely $j_\alpha^S=2\left(c_\alpha^++c_\alpha^-\right),	j_\alpha^P=-2\,\imu\left(c_\alpha^+-c_\alpha^-\right)$.

Next, we return to Newton's and Girard's formula, Eq.\,\eqref{newton}, and use
\begin{align*}
 	\text{tr}\left(\mathcal{J}^\pm\right)&=2\sqrt{\dfrac{3}{2}}c_0^\pm\\
	\text{tr}\left({\mathcal{J}^\pm}^2\right)&=\text{tr}\left(c_\alpha^\pm\lambda_\alpha c_\beta^\pm\lambda_\beta\right)=c_\alpha^\pm c_\beta^\pm 2\delta_{\alpha\beta}=2 c_\alpha^\pm c_\alpha^\pm\\
	\text{tr}\left({\mathcal{J}^\pm}^3\right)&=c_\alpha^\pm c_\beta^\pm c_\gamma^\pm\text{tr}\left(\lambda_\alpha\lambda_\beta\lambda_\gamma\right).
\end{align*}
Inserting this into Eq.\,\eqref{newton}, one has
\begin{align*}	\det\mathcal{J}^++\det\mathcal{J}^-&=2\sqrt{\dfrac{3}{2}}\left[{c_0^+}^3+{c_0^-}^3\right]-2\sqrt{\dfrac{3}{2}}\left[c_0^+c_\alpha^+c_\alpha^++c_0^-c_\alpha^-c_\alpha^-\right]+\\
	&\quad+\left[c_\alpha^+c_\beta^+c_\gamma^++ c_\alpha^-c_\beta^-c_\gamma^-\right]\text{tr}\left(\lambda_\alpha\lambda_\beta\lambda_\gamma\right).
\end{align*}
Now, using the relations between the $c$'s and the currents $j^S,j^P$, we have
\begin{align*}
 	{c_0^+}^3+{c_0^-}^3=2\text{Re}\left[{c_0^+}^3\right]&=\dfrac{1}{32}j_0^S\left({j_0^S}^2-3 {j_0^P}^2\right)\\
	c_0^+c_\alpha^+c_\alpha^++c_0^-c_\alpha^-c_\alpha^-=2\text{Re}\left[c_0^+c_\alpha^+c_\alpha^+\right]&=\dfrac{1}{32}\left[j_0^S\left({j_\alpha^S}^2-{j_\alpha^P}^2\right)-2 j_0^Pj_\alpha^Sj_\alpha^P\right]\\
	c_\alpha^+c_\beta^+c_\gamma^++c_\alpha^-c_\beta^-c_\gamma^-=2\text{Re}\left[c_\alpha^+c_\beta^+c_\gamma^+\right]&=\dfrac{1}{32}\left[j_\alpha^S\left(j_\beta^Sj_\gamma^S-j_\beta^Pj_\gamma^P\right)-j_\alpha^P\left(j_\beta^Sj_\gamma^P+j_\gamma^Sj_\beta^P\right)\right].
\end{align*}
Inserting this in the previous formula leads to
\begin{align*}
 	\det\mathcal{J}^++\det\mathcal{J}^-&=\dfrac{1}{16}\sqrt{\dfrac{3}{2}}j_0^S\left({j_0^S}^2-3{j_0^P}^2\right)-\dfrac{1}{16}\sqrt{\dfrac{3}{2}}\left[j_0^S\left({j_\alpha^S}^2-{j_\alpha^P}^2\right)-2 j_\alpha^Sj_0^Pj_\alpha^P\right]+\\
&\quad+\dfrac{1}{96}\left[j_\alpha^Sj_\beta^Sj_\gamma^S-3j_\alpha^Sj_\beta^Pj_\gamma^P\right],
\end{align*}
where summation over $\alpha,\beta,\gamma\in\{0,\dots,8\}$ is implicit.

Finally, using the $\text{SU}(3)$ structure constants $f_{k\ell m},d_{k\ell m}$, defined through $[\lambda_k,\lambda_\ell]=2\,\imu\,f_{k\ell m}\lambda_m$ and $\{\lambda_k,\lambda_\ell\}=\frac{4}{3}\delta_{k\ell}+2 d_{k\ell m}\lambda_m$, respectively, one obtains $\lambda_k\lambda_\ell=\imu\,f_{k\ell m}\lambda_m+d_{k\ell m}\lambda_m+\frac{2}{3}\delta_{k\ell}$ and, hence,
\begin{equation}
	\text{tr}\left(\lambda_k\lambda_{\ell}\lambda_i\right)=2\,\imu\, f_{k{\ell} m}\delta_{m i}+2 d_{k{\ell} m}\delta_{m i}
\end{equation}
(for $k,\ell,m,i\in\{1,\dots,8\}$) which allows us to write
\begin{align*}
 	\det\mathcal{J}^++\det\mathcal{J}^-&=\dfrac{1}{48}\sqrt{\dfrac{2}{3}}\left({j_0^S}^3-3j_0^S{j_0^P}^2\right)-\dfrac{1}{32}\sqrt{\dfrac{2}{3}}\left(j_0^Sj_k^Sj_k^S-j_0^Sj_k^Pj_k^P-2j_k^Sj_0^Pj_k^P\right)+\\
	&\quad +\dfrac{1}{48} d_{\ell k m}\left(j_\ell^S j_k^S j_m^S-3j_\ell^Sj_k^Pj_m^P\right),
\end{align*}
where, again, $\ell,k,m\in\{1,\dots,8\}$.

If one sets
\begin{equation}
 	\mathcal{A}_{\alpha\beta\gamma}:=\dfrac{1}{3!}\varepsilon_{ijk}\varepsilon_{mn\ell}\left(\lambda_\alpha\right)_{im}\left(\lambda_\beta\right)_{jn}\left(\lambda_\gamma\right)_{kl}\qquad\text{for $\alpha,\beta,\gamma\in\{0,\dots,8\}$},
\end{equation}
then the  expression above can be written in a more compact form as
\begin{equation}
 	\det\mathcal{J}^++\det\mathcal{J}^-=\dfrac{1}{32}\mathcal{A}_{\alpha\beta\gamma}\left(j_\alpha^Sj_\beta^Sj_\gamma^S-3j_\alpha^Sj_\beta^Pj_\gamma^P\right).
\end{equation}
This is the form given in Eq.~\eqref{sint6sep}, with the coupling constant properly adjusted.

\end{section}

\begin{section}{Second-order contributions to the action}\label{secondorderapp}

We demonstrate how the expressions for the inverse meson propagators, $G_P$, can be derived. The self-energy contribution $\varPi_{ij}$ can easily be derived from the fermion determinant, Eq.~\eqref{fermdet}, using the following formulas for the functional derivatives
\begin{equation*}
 	\begin{aligned}	\dfrac{\delta}{\delta\pi_{ij}(k)}\ln\,\det\hat{\mathscr{A}}&=\dfrac{\delta}{\delta\pi_{ij}(k)}\text{Tr}\,\ln\hat{\mathscr{A}}=\text{Tr}\left\{\hat{\mathscr{A}}^{-1}(p,p')\dfrac{\delta\hat{\mathscr{A}}(p',p'')}{\delta\pi_{ij}(k)}\right\}\\	\dfrac{\delta}{\delta\pi_{ij}(k)}\hat{\mathscr{A}}^{-1}&=-\hat{\mathscr{A}}\dfrac{\delta\hat{\mathscr{A}}}{\delta\pi_{ij}}\hat{\mathscr{A}}^{-1}.
\end{aligned}
\end{equation*}
 	
Moreover,  owing to the SPA equations, Eq.\,\eqref{spa}, the auxiliary fields\footnote{from now on we omit the tildes on $\tilde S_\alpha,\tilde P_\alpha$} $S_\alpha,P_\alpha$  are implicit functions of  $\sigma$ and $\pi$. This implies calculating the second derivative of the expression
\begin{equation*}
 	\tilde{\mathcal{S}}_\text{E}:=\sigma_\alpha S_\alpha+\pi_\alpha P_\alpha+\dfrac{G}{2}\left(S_\alpha S_\alpha+P_\alpha P_\alpha\right)+\dfrac{H}{4}\mathcal{A}_{\alpha\beta\gamma}\left(S_\alpha S_\beta S_\gamma-3 S_\alpha P_\beta P_\gamma\right)\,.
\end{equation*}
Neglecting first the space dependence of the fields we may first introduce the matrices $\overleftrightarrow{\sigma}=\frac{1}{\sqrt{2}}\sigma_\alpha\lambda_\alpha$ and $\overleftrightarrow{\pi}=\frac{1}{\sqrt{2}}\pi_\alpha\lambda_\alpha$.  The SPA equations \eqref{spa} in this new basis then read
\begin{subequations}
	\begin{align}
 	 	\sqrt{2}\overleftrightarrow{\sigma}+G S_\alpha\lambda_\alpha+\dfrac{3H}{4}\mathcal{A}_{\alpha\beta\gamma}\lambda_\alpha(S_\beta S_\gamma-P_\beta P_\gamma)&=0 \label{spa1}\\
		\sqrt{2}\overleftrightarrow{\pi}+G P_\alpha\lambda_\alpha-\dfrac{3 H}{2}\mathcal{A}_{\alpha\beta\gamma}\lambda_\alpha S_\beta P_\gamma&=0.\label{spa2}
	\end{align}
\end{subequations}
From the first derivative of Eq.~\eqref{spa1},
\begin{equation*}
 	0+G\dfrac{\delta S_\alpha}{\delta \pi_{ij}}\lambda_\alpha+\dfrac{3H}{4}\mathcal{A}_{\alpha\beta\gamma}\lambda_\alpha\left(2 S_\beta\dfrac{\delta S_\gamma}{\delta \pi_{ij}}-2 P_\beta\dfrac{\delta P_\gamma}{\delta\pi_{ij}}\right)=0,
\end{equation*}
it follows that $\frac{\delta S_\alpha}{\delta\pi_{ij}}=0$ for all $\alpha\in\{0,\dots,8\}$ and $i,j\in\{1,2,3\}$, recalling that $P_\alpha=0$ for all $\alpha$ in mean-field approximation. \\
The second derivative of Eq.~\eqref{spa1} leads to
\begin{equation}\label{ii}
 	G\lambda_\alpha\dfrac{\delta^2 S_\alpha}{\delta\pi_{k\ell}\,\delta\pi_{ij}}+\dfrac{3H}{2}\mathcal{A}_{\alpha\beta\gamma}\lambda_\alpha S_\beta\dfrac{\delta^2 S_\gamma}{\delta\pi_{k\ell}\,\delta\pi_{ij}}=\dfrac{3 H}{2}\mathcal{A}_{\alpha\beta\gamma}\lambda_\alpha \dfrac{\delta P_\beta}{\delta\pi_{k\ell}}\dfrac{\delta P_\gamma}{\delta \pi_{ij}}.
\end{equation}
Analogously, one has from the second equation
\begin{equation*}
 	\sqrt{2}\delta_{im}\delta_{jn}+G\dfrac{\delta P_\alpha}{\delta\pi_{ij}}\left(\lambda_\alpha\right)_{mn}-\dfrac{3H}{2}\mathcal{A}_{\alpha\beta\gamma}\left(\lambda_\alpha\right)_{mn}S_\beta\dfrac{\delta P_\gamma}{\delta\pi_{ij}}=0
\end{equation*}
or, by contraction with $\left(\lambda_\epsilon\right)_{nm}$
\begin{equation}\label{iii}
	G\dfrac{\delta P_\epsilon}{\delta\pi_{ij}}-\dfrac{3H}{2}\mathcal{A}_{\epsilon\beta\gamma} S_\beta\dfrac{\delta P_\gamma}{\delta\pi_{ij}}=-\dfrac{1}{\sqrt{2}}\left(\lambda_\epsilon\right)_{ij}\,.
\end{equation}
Finally, from the second derivative
\begin{equation*}
	0+G\dfrac{\delta^2 P_\alpha}{\delta\pi_{k\ell}\,\delta\pi_{ij}}\left(\lambda_\alpha\right)_{mn}-\dfrac{3 H}{2}\mathcal{A}_{\alpha\beta\gamma}\left(\lambda_\alpha\right)_{mn} S_\beta\dfrac{\delta^2 P_\gamma}{\delta\pi_{k\ell}\,\delta\pi_{ij}}=0\,,
\end{equation*}
and it follows that $\frac{\delta^2 P_\alpha}{\delta\pi_{k\ell}\,\delta\pi_{ij}}=0$ for all $\alpha,i,j,k,\ell$ in mean-field approximation.\\
The sum of the SPA equations gives
\begin{equation*}
 	\sigma_\alpha S_\alpha+\pi_\alpha P_\alpha+G(S_\alpha S_\alpha+P_\alpha P_\alpha)+\dfrac{3H}{4}\mathcal{A}_{\alpha\beta\gamma}(S_\alpha S_\beta S_\gamma-3 S_\alpha P_\beta P_\gamma)=0\,,
\end{equation*}
so that one can write
\begin{equation*}
 	\tilde{\mathcal{S}}_\text{E}=-\dfrac{1}{2} G(S_\alpha S_\alpha +P_\alpha P_\alpha)-\dfrac{H}{2}\mathcal{A}_{\alpha\beta\gamma}(S_\alpha S_\beta S_\gamma-3 S_\alpha P_\beta P_\gamma)\,.
\end{equation*}
Finally, applying identities \eqref{ii} and \eqref{iii} we may deduce the desired derivative 
\begin{equation*}	\dfrac{\delta^2\tilde{\mathcal{S}}_\text{E}}{\delta\pi_{k\ell}\,\delta\pi_{ij}}=\dfrac{1}{\sqrt{2}}\left(\lambda_\beta\right)_{ij}\dfrac{\delta P_\beta}{\delta\pi_{k\ell}}\,.
\end{equation*}
We conclude that the additional term is given by the solution of relation \eqref{iii} contracted by $\lambda_\alpha$,
\begin{equation*}
 	G\dfrac{\delta P_\alpha}{\delta\pi_{ij}}\left(\lambda_\alpha\right)_{mn}-\dfrac{3H}{2}\mathcal{A}_{\alpha\beta\gamma}
\left(\lambda_\alpha\right)_{mn} S_\beta\dfrac{\delta P_\gamma}{\delta\pi_{ij}}=
-\sqrt{2}\delta_{i m}\delta_{jn}\,;
\end{equation*}
this can be further simplified by noting $S_\alpha=\frac{1}{2}\text{tr}(\lambda_\alpha S)$ and
\begin{align*}
 	\mathcal{A}_{\alpha\beta\gamma}\left(\lambda_\alpha\right)_{mn} S_\beta\dfrac{\delta P_\gamma}{\delta\pi_{ij}}&=\dfrac{1}{3!}\varepsilon_{rsk}\varepsilon_{uv\ell}\left(\lambda_\alpha\right)_{ru}\left(\lambda_\beta\right)_{sv}\left(\lambda_\gamma\right)_{k\ell}\dfrac{1}{2} S_t\left(\lambda_\beta\right)_{tt}\left(\lambda_\alpha\right)_{mn}\dfrac{\delta P_\gamma}{\delta\pi_{ij}}\\
&=\dfrac{1}{3}\varepsilon_{ntk}\varepsilon_{mt\ell}\dfrac{\delta P_\gamma}{\delta\pi_{ij}}\left(\lambda_\gamma\right)_{k\ell} S_t\,.
\end{align*}
Consequently, the equation to be solved is
\begin{equation*}
 	G\left(\lambda_\alpha\right)_{mn}\dfrac{\delta P_\alpha}{\delta\pi_{ij}}-\dfrac{H}{2}\varepsilon_{kn t}\varepsilon_{t\ell m} S_t\left(\lambda_\gamma\right)_{k\ell}\dfrac{\delta P_\gamma}{\delta\pi_{ij}}=-\sqrt{2}\delta_{im}\delta_{jn}.
\end{equation*}
Defining $\left(r_{ij,mn}\right)^{-1}:=\frac{1}{\sqrt{2}}\left(\lambda_\alpha\right)_{mn}\frac{\delta P_\alpha}{\delta\pi_{ij}}$ we may write
\begin{equation}
 	\dfrac{\delta^2\tilde{\mathcal{S}}_\text{E}}{\delta\pi_{k\ell}\,\delta\pi_{ij}}=-\left(r_{ij,k\ell}\right)^{-1},
\end{equation}
where $r_{ij,k\ell}$ solves the system given in Eq.~\eqref{rsystem}.

Finally, we consider the functional derivative of terms of the form $\int\diff^4 x\,S_\alpha(x) S_\beta(x) S_\gamma(x)$ etc. The first derivative with respect to $\pi_{ij}(y)$ generates a $\delta$ function, $\delta(x-y)$, hence 
\begin{equation*}
	\int\diff^4 x\,S_\alpha(x)S_\beta(x)S_\gamma(x)\to S_\alpha(y)S_\beta(y)S_\gamma(y)\,.
\end{equation*}
 The second derivative with respect to $\pi_{k\ell}(z)$ generates an additional $\delta(y-z)$. This means that in mean-field approximation the functional dependence of the fields after a Fourier transformation is given by
\begin{equation*}
 	r_{ij,k\ell}^{-1}\int\diff^4y\,\diff^4 z\,\euler^{-\imu p\cdot y}\,\euler^{-\imu p'\cdot z}\,\delta(y-z)\,\delta\pi_{ij}(y)\,\delta\pi_{k\ell}(z)=r_{ij,k\ell}^{-1}\,\delta\pi_{ij}(p)\,\delta\pi_{k\ell}(-p).
\end{equation*}
Treating  analogously  the contributions  from the $\sigma$ field, we arrive at Eq.~\eqref{se2}.

\end{section}

\end{appendix}

\end{document}